\documentclass[tightenlines,eqsecnum,floats,aps,amsmath,nofootinbib,amssymb,prd,shownopacs,notitlepage]{revtex4-1}
\usepackage{epsf}

\usepackage{graphicx, wrapfig}
\usepackage{epstopdf}
\usepackage{amsthm}
   \usepackage{amsfonts}
\usepackage{amssymb}
\usepackage{color}
\usepackage{amsmath}
\usepackage{subfigure}
\makeatletter
\def\hlinewd#1{%
  \noalign{\ifnum0=`}\fi\hrule \@height #1 \futurelet
   \reserved@a\@xhline}
\makeatother

\DeclareGraphicsExtensions{.eps,.jpg,.png,.pdf}

\def\f{\frac}

\def\d{\textrm{d}}

\def\lp{l_{\rm Pl}}

\usepackage{enumerate}

\newcommand{\be}{\nopagebreak[3]\begin{equation}}
\newcommand{\ee}{\end{equation}}
\newcommand{\bfig}{\nopagebreak[3]\begin{figure}}
\newcommand{\efig}{\end{figure}}
\newcommand{\ba}{\nopagebreak[3]\begin{eqnarray}}
\newcommand{\ea}{\end{eqnarray}}

\newcommand{\bmult}{\nopagebreak[3]\begin{multline}}
\newcommand{\emult}{\end{multline}}

\newcommand{\eref}[1]{eq.\,(\ref{#1})}

\newcommand{\code}[1]{{\tt #1}}

\begin{document}
\title[Numerical simulations of loop quantum Bianchi-I spacetimes]{Numerical simulations of loop quantum Bianchi-I spacetimes}

\author{Peter Diener$^{1,2}$}
\email{diener@cct.lsu.edu}

\author{Anton Joe$^2$}
\email{Deceased}

\author{Miguel Megevand$^{2,3}$}
\email{megevand@phys.lsu.edu}

\author{Parampreet Singh$^{2,1}$} 
\email{psingh@phys.lsu.edu}

\affiliation{
$^1$ 
Center for Computation and Technology, \\ Louisiana State University, Baton Rouge, LA 70803, U.S.A.
}
\affiliation{
$^2$ 
Department of Physics and Astronomy, \\ Louisiana State University, Baton Rouge, LA 70803, U.S.A.
}
\affiliation{$^3$ Instituto de F\'isica Enrique Gaviola, CONICET \\
C\'ordoba, Argentina}

\begin{abstract}
Due to the numerical complexities of studying evolution in an anisotropic quantum spacetime, in comparison to the isotropic models, the 
physics of loop quantized anisotropic models has remained largely unexplored.  In particular, robustness of bounce  and the validity of effective dynamics have so far not been established. Our analysis fills these gaps for the case of vacuum Bianchi-I 
spacetime. To efficiently solve the quantum Hamiltonian constraint we perform an implementation of the Cactus framework which is 
conventionally used for applications in numerical 
relativity. Using high performance computing, numerical simulations for a large number of initial states with a wide variety of fluctuations 
are performed. Big bang singularity is found to be replaced by anisotropic bounces  for all the cases. We find that for initial 
states which are sharply peaked at the late times in the classical regime and bounce at a mean volume much greater than the Planck volume, effective dynamics is an 
excellent approximation to the underlying quantum dynamics. Departures of the effective dynamics from the quantum evolution appear for the 
states probing deep Planck volumes. A detailed analysis of the behavior of this departure reveals a non-monotonic and subtle dependence on 
fluctuations of the initial states. We find that effective dynamics in almost all of the cases underestimates the volume and hence overestimates the curvature at the bounce, a result in synergy with earlier 
findings in isotropic case. The expansion and shear scalars are found to be bounded throughout the evolution. 
\end{abstract}

\maketitle

\section{Introduction}\label{sec:intro}

In order to understand the generic approach to the classical singularities and their resolution, the role of anisotropic spacetimes 
is fundamental. Of these the Bianchi-I spacetime is one of the simplest yet an important one to study. In the classical theory, 
rigorous analytical and numerical techniques have established that in cosmological spacetimes the singularity either has a Kasner form 
corresponding to the vacuum Bianchi-I spacetime or undergoes a Mixmaster dynamics which is made of a sequence of Kasner 
phases \cite{berger,garfinkle}. For homogeneous cosmological spacetimes without spatial curvature and even 
with a tiny initial anisotropy, the singularity structure is determined by the anisotropic shear unless the 
matter has stiff or ultra-stiff 
equation of state. For any other equation of state, the metric near the singularity is guaranteed to correspond to the vacuum Bianchi-I 
metric. The presence of anisotropies brings considerable richness and complexity 
to the gravitational dynamics. As an example, it changes the point type big bang singularity 
to a cigar shaped big bang. As this cigar singularity is approached, two of the three directional scale factors 
contract to zero in a finite proper time. The expansion and shear scalars diverge, causing a divergence in the Weyl curvature as well as the 
Ricci curvature (if matter is present). It has been hoped that insights on the quantum nature of spacetime would provide answers to how to 
the spacetime extends beyond these singularities. Certainly one of the most formidable challenge for any theory of quantum gravity is 
whether such singularities can be resolved. 

The issue of singularity resolution has been rigorously addressed in a non-perturbative approach to the quantization of homogeneous cosmological spacetimes 
known as loop quantum cosmology (LQC) \cite{as-status,as-rev}. In this framework based on loop quantum gravity, loop quantization of various 
isotropic and anisotropic spacetimes has been performed in the last decade. The key result, which was first shown for the case of a 
spatially flay homogeneous and isotropic model with a massless scalar field, is that the big bang singularity is resolved due to the discrete 
quantum geometric effects near the Planck scale and is replaced by a big bounce \cite{aps1,aps2,aps3}. The existence of bounce first 
established using numerical simulations was confirmed via an exactly solvable model  which predicts a minimum volume for all the states in 
the physical Hilbert space \cite{slqc} and consistent probability for singularity resolution to be unity \cite{consistent-lqc}. In the 
isotropic models the results on singularity resolution have been 
generalized to include spatial curvature \cite{apsv,closed-warsaw,kv-open}, radiation \cite{rad}, 
and cosmological constant \cite{bp-lambda,kp-lambda,ap-lambda}. In all these models,  
quantum expectation values of the relational observables show that at small spacetime curvature there is an excellent 
agreement between the quantum dynamics of LQC and general relativity (GR). When the spacetime curvatures reach about a percent of the Planck 
value, strong departures start becoming significant  between the two. Unlike a contracting universe ending up in a big bang in the backward 
evolution, the 
LQC universe bounces when the expansion scalar 
of the isotropic spacetime reaches a universal maximum \cite{cs-geom, ps09}.\footnote{Conventionally, bounce in isotropic models 
has been characterized by a universal maximum in energy density, see for eg. \cite{ps06,aps3}. However, it has been recently shown that a 
more faithful characterization of the bounce in LQC which is valid irrespective of spatial curvature is expansion scalar \cite{ds16}. In the 
vacuum Bianchi-I spacetime as considered here, one of our goals will be to understand the behavior of expansion and shear scalars.} 

In isotropic models in LQC, numerical simulations have played a crucial role in deciphering the details of singularity resolution and 
associated Planck scale physics \cite{ps12,khanna-review}. A key difference in the numerical techniques used in LQC in comparison to 
numerical relativity concerns with the nature of  evolution equation. Unlike in GR where the spacetime is a continuous differentiable 
manifold and the evolution equations are differential equations, in LQC the evolution equation in the geometric representation is 
fundamentally discrete. There is no freedom to change the discreteness to achieve a stable evolution. This puts non-trivial 
computational constraints, especially if one wishes to study fate of singularity resolution for widely spread states, and anisotropic spacetimes. In the 
former case, if one is interested in simulations involving general states then recently proposed Chimera scheme proves quite useful 
\cite{numlsu-1}. Using this scheme, singularity resolution has been 
established for a wide variety of 
states and the validity of an effective spacetime description in LQC has been rigorously studied. The latter description which captures the 
LQC dynamics quite well in a continuum spacetime has proved extremely valuable to extract physical implications and phenomenological 
predictions \cite{as-status,as-rev}. It was recently found that though effective description is an excellent approximation for states which 
are initially sharply peaked in a macroscopic universe and bounce at volumes large compared to Planck volume, the situation changes if 
states are widely spread and probe deep Planckian geometry \cite{numlsu-2,numlsu-3}. In this case effective dynamics underestimates the volume 
at the bounce, thus overestimating the spacetime curvature where singularity resolution occurs. Further, the dependence of the departure 
between the quantum and effective dynamics was found to non-trivially and unexpectedly non-monotonically depend on the state parameters and fluctuations. 

Unlike the case of isotropic models where there exist thorough studies on the fate of singularities in LQC, investigations on Bianchi models have been so far 
focused on two frontiers: (i) establishing the details of 
quantization 
\cite{martin-bianchi,chiou-bianchi1,b1-madrid1,awe-bianchi1,awe-bianchi2,we-bianchi9,pswe,corichi-bianchi3} and (ii) 
using effective dynamics to understand genericity of singularity resolution \cite{martin-shyam,cs-geom,ps11,ps-proc,corichi-bianchi1,ps16} 
and extract various phenomenological aspects (see for eg. 
\cite{martin-shyam-golam,chiou-bianchi3,kevin-chiou,magnetic,bgps-spatial,bgps-kasner,bgps-inflation,corichi-bianchi2}). In contrast 
there has been only one study on the numerical aspects using quantum states in Bianchi-I model  \cite{b1-madrid2}.\footnote{These spacetimes have also been discussed to understand numerical stability of quantum difference equation, see for eg. \cite{martin-cartin-khanna,marie-bianchi}.} This 
seminal study which established a quantization prescription for Bianchi-I spacetime in LQC demonstrated that  bounce occurs at least for  sharply peaked Gaussian state. In this first important step to understand singularity resolution,  the details of the validity of effective dynamics and the robustness of bounce and associated features were not studied. 
Without a rigorous numerical analysis with sharply peaked as well as widely spread states, robustness of singularity resolution and bounce in the loop quantization of Bianchi-I spacetime and 
the validity of effective dynamics can not be established. Despite the availability of a consistent quantization prescription where physical Hilbert space, inner product 
and relational observables are known, numerical analysis and hence analysis of physical implications of quantum theory had so far been been limited because of additional numerical complexities and very high computational costs. 

The goal of this manuscript is to fill the above important gap in the understanding of Bianchi-I spacetimes 
in LQC. We focus on the quantization prescription put forward in 
Refs. \cite{b1-madrid1,b1-madrid2} which provides a viable quantization when the spatial manifold is a 3-torus. Note that the quantization 
prescription which we employ in this analysis is not a unique choice to loop quantize Bianchi-I spacetimes. There exists another loop 
quantization, first proposed in Ref. \cite{chiou-bianchi1} and rigorously studied in Ref. \cite{awe-bianchi1}. 
These choices result from  quantization ambiguities which arise due to different prescriptions to obtain the field strength of the Ashtekar-Barbero 
connections and the way loops over which holonomies of these connection are considered capture the underlying quantum geometry. Surprisingly 
these are the only two viable choices if the spatial manifold is compact \cite{cs-geom}.   If the spatial manifold is non-compact, the  
quantization prescription of Refs. \cite{b1-madrid1,b1-madrid2} results in physics which depends on certain rescalings of the shape of the 
fiducial cell used to define the symplectic structure \cite{cs-geom}.  In comparison, 
the quantization prescription developed in Refs. \cite{awe-bianchi1} promises to provide a viable quantization irrespective of the choice 
of the topology of the spatial manifold. However, so far some of the important 
details of the physical Hilbert space which are crucial to perform analytical and numerical investigations have remained unavailable in the latter approach. On the other hand, though the 
quantization prescription of Refs. \cite{b1-madrid1,b1-madrid2} is restricted to a 3-torus spatial topology, details of the physical Hilbert 
space are rigorously available. Finally, both the viable quantizations result in the so called improved dynamics prescription in the 
isotropic limit \cite{cs-geom}, which by itself is a unique consistent choice for the loop quantization of isotropic models 
\cite{cs-unique}. It is to be noted that working with 3-torus topology should not be viewed as a restriction since the same setting is 
very useful to implement the loop quantization of polarized Gowdy models and understand the role of inhomogeneities on bounce  using a hybrid quantization \cite{hybrid-bounce} (see Ref. \cite{hybrid-rev} for a review).

As we will show later, numerical simulations of Bianchi-I spacetime in LQC for states which are widely spread and probe deep Planck regime 
come with significant computational challenges. In particular, we  need high performance computing (HPC) resources
to solve this problem and hence a parallelization of code is a necessity. We therefore 
employ the \code{Cactus} computational toolkit~\cite{Allen99a,cactus_web}.
\code{Cactus} was originally developed for numerical relativity in order
to solve the partial differential equations originating from  the full classical
general relativistic field equations. The design of \code{Cactus} is modular
and allows domain specialists to focus on their particular domain of expertise
while still writing inter-operable modules. Thus computer scientists can focus
on infrastructure for parallelization, I/O and other necessary things, while
the physicist can focus on the physics part of the implementation assuming some of the more technical aspects of the parallelization. In this way
\code{Cactus} provides an abstraction of parallelization that enables much
faster development of parallel codes. Once the \code{Cactus} implementation was performed for the loop quantum Bianchi-I spacetime, HPC 
resources of Extreme Science and Engineering Discovery Environment (XSEDE) \cite{xsede}
were used for our investigations in this manuscript. It is to be noted that our work is the first of its kind to use the \code{Cactus} framework and HPC in quantum gravity. 

Numerical simulations carried out in our analysis reveal various so far not known features of the Bianchi-I spacetime in 
LQC. We first establish rigorously the 
existence of bounce(s) of directional volumes for sharply peaked and widely spread Gaussian states. In over a hundred simulations carried out with different states, big bang 
singularity is found to be resolved and replaced by the bounce of mean volume. We find that for states which bounce at volumes much larger 
than the Planck volume, 
the effective spacetime description provides an excellent approximation to the underlying quantum dynamics. However, for states which probe the 
deeper Planck volumes there are departures of the effective dynamics from the quantum theory. It is important to note that in this manuscript 
we do not include any state dependent corrections to the effective Hamiltonian. Any reference to the departure of effective dynamics from 
the quantum theory will be made under this assumption. Having gained 
the evidence of existence of departure between the effective dynamics and quantum theory which turns out to be maximum in the bounce regime, 
we explore the dependence of the departure on state parameters $\omega_i$ whose inter-relationship captures  anisotropy and which are 
proportional to the directional Hubble rates in the classical theory, and fluctuations of the initial state. The Hamiltonian constraint relates 
three $\omega_i$'s, allowing only two of them say, $\omega_2$ and $\omega_3$, to be independent. Working in the mixed representation with 
conjugate of directional volume $v_1$ as time, the initial states are parameterized by $\omega_2$ and $\omega_3$. The measure of the 
departure between the quantum theory and the effective dynamics is found using the expectation value of the relational observable for 
logarithm of directional volume in the quantum theory and its analog in the effective dynamics when bounce occurs. We find that the 
departure of the effective dynamics from the quantum theory increases rapidly when one of the $\omega_i$ is decreased keeping the other one fixed.  
Varying the value of $\omega_i$ where  the state is peaked shows that the behavior of the departure is sensitive to the choice of a given dispersion 
for different values of $\omega_i$. There is an increase in the departure as one of the $\omega_i$  is decreased keeping the other one fixed, and 
some evidence of a non-monotonic 
behavior for larger values of $\omega_i$. 
The dependence of the departure of the effective dynamics from the quantum theory shows a striking non-monotonic behavior with respect to the 
dispersion in the logarithm of the directional volume. The departure takes largest values when the dispersion in logarithm of the directional volume is largest, 
however it does not take smaller values at the smallest values of this dispersion. There exists an unambiguous regime where the departures increase 
on decreasing the dispersion. We find that dispersion in logarithm of volume can only be decreased to a certain value for a given value of 
$\omega_i$. Our results show that departure of the effective dynamics for quantum theory has non-trivial and subtle  relationship with state 
parameters and fluctuations. We find that in most cases the states which have wide spreads bounce at volumes greater than those predicted by 
effective dynamics. That means, effective dynamics generally overestimates the spacetime curvature at the singularity resolution. This 
result is in harmony with the similar results earlier obtained for isotropic models in LQC \cite{numlsu-2,numlsu-3}. Interestingly, we do 
find some states 
for certain values of parameters and dispersions for which the difference in the expectation value of logarithm of the 
directional volume and its counterpart in the effective dynamics is negative. This means that there are certain states for which the above 
conclusion is reversed. Finally, using the expectation values of the relational observables and some inputs from the effective dynamics we 
estimate the mean Hubble rate (expansion scalar) and shear scalar for sharply peaked states. Along with the directional Hubble rates, these scalars turn 
out to be bounded.


This manuscript is organized as follows. In Sec. II we discuss the main features of the quantization of 
the Bianchi-I spacetime in absence of matter. We will follow the quantization 
methods rigorously outlined in Refs. \cite{b1-madrid1,b1-madrid2}, where various properties of the quantum Hamiltonian 
constraint were established and singularity resolution was shown for the first time at the quantum level. This section provides a summary of the quantization procedure, and for details the reader is referred to above references.  
In Sec. III we discuss various computational challenges, the ways to overcome them through {\code{Cactus}} implementation 
and the performance and scaling of 
our code. Readers who are mainly interested in the quantization and physical implications can skip this section and move to Sec. IV which deals with 
a detailed discussion of all the results from our analysis. We first discuss resolution of classical singularity by bounces in Sec. IVA, 
where we 
also discuss the way effective dynamics provides an excellent agreement with the quantum dynamics for sharply peaked states. This is 
followed by a discussion of 
some of the cases which show the departure of effective dynamics from the quantum theory in Sec. IVB. In Sec. IVC various quantitative details of this departure are investigated. 
In Sec. IVD we estimate the expansion and shear scalars, and deceleration parameter in our analysis. The mean Hubble rate (or the expansion scalar) and the shear scalar 
turn out to be bounded. We summarize our results with a summary and a discussion of open issues in Sec. V.

\section{Loop quantization of Bianchi-I vacuum spacetime}\label{sec:lqc}
 
We consider the spatial manifold with a 3-torus topology  which allows the 
quantization prescription put forward in Refs. \cite{b1-madrid1,b1-madrid2} to be successfully carried out.\footnote{If the spatial manifold is chosen as $\mathbb{R}^3$ then the quantization prescription is not independent of the 
choice of the fiducial cell \cite{awe-bianchi1,cs-geom}.} The $\mathbb{T}^3$ 
fiducial cell is chosen with sides of coordinate length $2 \pi$. Due to the underlying homogeneity of the 
spatial manifold, the matrix valued connection $A_i^a$ and triad variables $E^a_i$ can be written as a 
homogeneous pair $(c^i,p_i)$ satisfying $\{c^i,p_j\} = 8 \pi G \gamma \delta^i_j$. Here $\gamma \approx 0.2375$ 
is the Barbero-Immirzi parameter. The spacetime metric is given by 
\be
\d s^2 = -N^2 \d t^2 + a_i^2 \d x_i^2
\ee
where directional scale factors $a_i$ are kinematically related to the triads as $p_i = 4 \pi^2 a_j a_k \mathrm{sgn}(a_i a_j a_k)$ where $i\neq j \neq k$ can take values $1..3$. 
On the other hand, connection components are related to time derivatives of the 
scale factors dynamically which can be found from the Hamiltonian constraint which is the only non-trivial constraint in this setting due to the underlying symmetries. The classical Hamiltonian constraint is given by
\be
{\mathcal{\tilde C}}_{\mathrm{cl}} = -\frac{2}{\gamma^2 V} \left(c^1 p_1 \, c^2 p_2 + c^2 p_2 \, c^3 p_3 + c^3 p_3 \, c^1 p_1\right) \approx 0 \label{cl-C}
\ee
where $V$ denotes the physical volume of $\mathbb{T}^3$ cell: $V = |p_1 p_2 p_3|^{1/2}$.

On quantization, the basic kinematical variables in the loop quantization are the holonomies of connection components along the edges of the fiducial cell, 
and the fluxes of the triads which turn out to be proportional to the triads themselves. Elements of holonomies, $N_\mu = \exp(i \mu_i c^i/2)$, 
form an algebra of almost periodic functions which forms the configuration algebra whose completion with respect to the inner product $\langle \mu_i| \mu_i' \rangle = \delta_{\mu_i, \mu_i'}$ 
yields the kinematical Hilbert space ${\cal H}_{\mathrm{kin}}$. The eigenstates of the triad operators are given by $|\mu_i\rangle$ such that 
\be
\hat p_i |\mu_i\rangle = 4 \pi \gamma \lp^2 \mu_i |\mu_i\rangle ~.
\ee
The action of $\hat N_\mu$ on the eigenstates of triad operators is translational: ${\hat N_{\mu'_i}} |\mu_i\rangle = |\mu_i + {\mu'_i}\rangle$.
Though these elements of holonomies yield a translations which have uniform spacing in 
triad eigenvalues, the same does not hold true when one takes into account holonomies are considered along physical lengths which have inbuilt triad dependence. 
In particular, the physical length $\bar \mu_i \sqrt{|p_i|}$ is determined by the underlying quantum geometry as: ${\bar \mu_i}^2 |p_i| = \lambda^2$,
where $\lambda^2 = 4 \sqrt{3} \pi \gamma \lp^2$ is the minimum eigenvalue of the area operator computed in lop quantum gravity.  The above relationship complicates the action of the operator ${\hat N_{\bar\mu_i}}$ on triad eigenstates. Nevertheless changing the representation to (dimensionless) directional volume $v_i = 2/\sqrt{3} \lambda^{-3} |p_i|^{3/2}$ results in a uniform translation for holonomies considered over $\bar \mu$. That is,  
\be
\hat N_{\bar\mu_i} |v_i\rangle = |v_i + 1 \rangle ~.
\ee

Using the action of these operators, one can obtain the quantum Hamiltonian constraint corresponding to eq.(\ref{cl-C}) which turns out to be non-singular and which has the property that the zero volume state is decoupled \cite{b1-madrid1}.
 Using this property, it turns out to be more convenient to work with the densitized quantum Hamiltonian constraint which can be written as \cite{b1-madrid1,b1-madrid2}:
\begin{equation}\label{C-q}
\hat{\mathcal{C}}= -\frac{2}{\gamma^2}\left( \hat{\Theta}_1\hat{\Theta}_2 +
\hat{\Theta}_1\hat{\Theta}_3+\hat{\Theta}_2\hat{\Theta}_3\right) ~.
\end{equation}
Here $\hat{\Theta}_i$ are symmetric operators which act on corresponding basis states $|v_i\rangle$ as 
\begin{equation}
\hat{\Theta}_i|v_i\rangle=-i \frac{\lambda^2}{2\sqrt{3}}\left( f_+(v_i)|v_i+2\rangle-f_-(v_i)|v_i-2\rangle \right),
\end{equation}
where
\be
f_\pm(v_i)=g(v_i\pm 2) s_{\pm}(v_i) g(v_i)
\ee
with
\be
s_{\pm}(v_i)= \mathrm{sgn}(v_i\pm 2) + \mathrm{sgn}(v_i)
\ee
and
\be
g(v_i)= \begin{cases}
             \left| \left|1+\frac{1}{v_i}\right|^{1/3} 
             \left|1-\frac{1}{v_i}\right|^{1/3}\right|^{-1/2} \quad &{\rm
               if} \quad v_i\neq 0, \\
             0 \quad &{\rm if}\quad v_i=0.
        \end{cases}
\ee

We can see
that the action of the operators $\hat{\Theta}_i$ only connects states with
the label $v_i$ separated by two. Furthermore, the positive and negative
regions are also disconnected from each other. That is, $\hat{\Theta}_i$ only
connects states with $v_i$ in the semi lattices
\begin{equation}
\mathcal{L}^{\pm}_{\epsilon_i}=\{\pm(\epsilon_i+2n),\; n=0,1,2,... \},
\end{equation}
where $0<\epsilon_i\leq2$. The Hilbert subspaces
$\mathcal{H}^{\pm}_{\epsilon_i}$, defined as the Cauchy completions with
respect to the discrete inner product of the spaces spanned by $|v_i\rangle$ with
$v_i$ in each $\mathcal{L}^{\pm}_{\epsilon_i}$, are left invariant by the action
of $\hat{\Theta}_i$. We can then restrict the study to one particular Hilbert
space, say $\mathcal{H}^{+}_{\epsilon_1}\otimes \mathcal{H}^{+}_{\epsilon_2}
\otimes \mathcal{H}^{+}_{\epsilon_3}$.

The spectrum of the essentially self-adjoint operator $\hat{\Theta}_i$ is continuous \cite{b1-madrid1}. The eigenstates, with eigenvalue denoted by $\omega_i$, can be obtained explicitly from the recursive relations:
\begin{equation}
e^{\epsilon_i}_{\omega_i}(2+\epsilon_i) = 
-i\frac{\sqrt{3}\omega_i}{\Delta} \frac{e^{\epsilon_i}_{\omega_i}(\epsilon_i)}{g(2+\epsilon_i)g(\epsilon_i)}, 
\end{equation}
and
\begin{equation}
e^{\epsilon_i}_{\omega_i}(2n+2+\epsilon_i) = 
\frac{g(2n-2+\epsilon_i)}{g(2n+2+\epsilon_i)}
e^{\epsilon_i}_{\omega_i}(2n-2+\epsilon_i) -i\frac{\sqrt{3}\omega_i}{\Delta}
\frac{e^{\epsilon_i}_{\omega_i}(2n+\epsilon_i)}{g(2n+2+\epsilon_i)g(2n+\epsilon_i)}
\quad (n>0) .
\end{equation}
They are thus completely determined by their value at the minimum allowed value. 
Note that these eigenfunctions, which have support in
$\mathcal{L}^{\pm}_{\epsilon_i}$, are formed by two components, each with
support in the four-step lattices, given respectively by:
\be
{}^{(4)}\mathcal{L}^{\pm}_{\epsilon_i}=\{\pm(\epsilon_i+4n),\; n=0,1,2,... \}, 
\ee
and
\be
{}^{(4)}\mathcal{L}^{\pm}_{\epsilon_i+2}=\{\pm(\epsilon_i+2+4n),\; n=0,1,2,... \}.
\ee
These components are eigenfunctions of $\hat{\Theta}_i^2$ with eigenvalue
$\omega_i^2$, and have a relative phase of $\pm \pi/2$. Choosing the initial
value $e^{\epsilon_i}_{\omega_i}(\epsilon_i)$ to be positive we obtain a
purely real and a purely imaginary component. In the numerical computation we
choose somehow random positive values and then eliminate this last degree of freedom
by rescaling the eigenfunction. The rescaling is chosen such that the
asymptotic forms match those of the Wheeler-DeWitt eigenfunctions.

States belonging to the physical Hilbert space ${\cal H}_{(\epsilon_1,\epsilon_2,\epsilon_3)}$, found using group averaging, can be written
 in terms of eigenfunctions $e_{\omega_i}^{\epsilon_i}(v_i)$ as:
\begin{equation}
\Psi(v_1,v_2,v_3)=\int d\omega_2 d\omega_3 \tilde{\Psi}(\omega_2,\omega_3) e^{\epsilon_1}_{\omega_1(\omega_2,\omega_3)}(v_1)
e^{\epsilon_2}_{\omega_2}(v_2) e^{\epsilon_3}_{\omega_3}(v_3), \\
\end{equation}
where $\tilde{\Psi}(\omega_2,\omega_3)$ is the wave profile, and 
we have chosen to write $\omega_1$ in terms of $\omega_2$ and $\omega_3$ as
\begin{equation}
\omega_1(\omega_2,\omega_3)= - \frac{\omega_2 \omega_3}{\omega_2+\omega_3} ~.
\end{equation}
This relationship is essential for the states to satisfy the quantum Hamiltonian constraint $\hat{\mathcal{C}} \Psi=0$. 

Once the physical Hilbert space is available, our next task is to extract relational dynamics. As an example, we can study the 
behavior of $v_2$ and $v_3$ in terms of $v_1$ or its conjugate variable $b_1$ which acts as an internal clock. It turns out that in the quantum theory, 
the choice of $v_1$ as internal time is unsuitable since it does not lead to a unitary evolution \cite{b1-madrid1}. It also turns out that 
 $v_1$ is not monotonous in the
proper time. On the other hand, unitary evolution can be found with $b_1$ as internal time. 
A physical state in the $b_i$ representation can be obtained from the one in the $v_i$ representation via the following Fourier 
transformation
\begin{equation} \label{eq:F}
\left[{\mathcal F}\Psi\right](b_i)=\sum_{v_i} \Psi(v_i) |v_i|^{-1/2}e^{-(i/2)v_ib_i}.
\end{equation}
In the following, we will work with a mixed representation, with wavefunctions depending on $b_1$, $v_2$ and $v_3$. In this case,  
the summation in~\eqref{eq:F}
should in principle be preformed in all of ${\mathcal L}^+_{\epsilon_1}$, with
$b_1$ going from $0$ to $2\pi$. 
However, as noted
in~\cite{b1-madrid1}, one can instead apply two separate transformations, each acting on
one of the sectors  $^{(4)}{\mathcal L}^+_{\epsilon_1}$ and $^{(4)}{\mathcal
  L}^+_{\epsilon_1+2}$, which map to $b_1$ in the domain $[0,\pi)$ and
  $[\pi,2\pi)$, respectively. Since all the information about the physical
    state is contained in each sector, we can restrict our analysis to only
    one of them. In what follows we will perform the transformations to $b_1$
    space as defined in~\eqref{eq:F}, but with the summation evaluated in $^{(4)}{\mathcal
      L}^+_{\epsilon_1}$ and hence with $b_1$ in the domain $[0,\pi)$.
To write the physical state in the mixed representation, we only need to transform the eigenstates in the $v_1$ direction:
\begin{equation} \label{eq:etilde}
{\tilde e}^{\epsilon_1}_{\omega_1}(b_1) :=\left[{\mathcal F} e^{\epsilon_1}_{\omega_1}\right](b_1)
= \sum_{^{(4)}{\mathcal
      L}^+_{\epsilon_1}} e^{\epsilon_1}_{\omega_1}(v_1) v_1^{-1/2}e^{-(i/2)v_ib_i}.
\end{equation}
The physical state in the mixed representation for evolution in $b_1$ can be written as:\footnote{To simplify the notation
  from now on we will drop the 
  the supra-index $\epsilon_i$. }
\be
\label{eq:chilqc}{ {\chi}_{b_1}}(v_2,v_3)=\int {\rm d}\omega_2 {\rm d}\omega_3 \tilde{ {\chi}}_{b_1}(\omega_2,\omega_3) e_{\omega_2}(v_2) e_{\omega_3}(v_3),
\ee
where 
\be
{ {\tilde \chi}}_{b_1}(\omega_2,\omega_3)=\tilde{  \Phi}(\omega_2,\omega_3) {\tilde e}^\prime_{\omega_1}(b_1)
\ee
 and 
 \begin{equation} \label{eprime}
 {\tilde e}^\prime_{\omega_1}(b)=\f{{\tilde e}_{\omega_1}(b)}{|{\tilde e}_{\omega_1}(b)|}.
 \end{equation}
The physical inner product evaluated at a `time' slice $b_1$ is given by
\be
\langle{  \chi}_{b_1}{  |\chi}_{b_1}^\prime\rangle=\sum_{v_2,\,v_3} \overline{  \chi}_{b_1}(v_2,v_3) {  \chi}_{b_1}^\prime(v_2,v_3) ~.
\ee

Obtaining the expectation values of the relational observables $\widehat{\ln(v_2)}|_{b_1}$ and $\widehat{\ln(v_3)}|_{b_1}$
is straightforward, as on a given slice $b_1$ they simply act as multiplication on the eigenstates
$e_{\omega_2}(v_2)$ and $e_{\omega_3}(v_3)$, respectively. For instance, we
have
\begin{equation}\label{lnv2}
\widehat{ \ln(v_2)_{b_1}}{ {\chi}_{b}}(v_2,v_3)=\int {\rm d}\omega_2 {\rm d}\omega_3 \tilde{ {\chi}}_{b}(\omega_2,\omega_3) \left[\ln(v_2)e_{\omega_2}(v_2)\right] e_{\omega_3}(v_3),
\end{equation}
and with similar expressions for $\widehat{\ln(v_3)}|_{b_1}$ and their dispersions.

On the other hand, the evaluation of the expectation value of ${\widehat{\ln(v_1)_{b_1}}}$ is
slightly more complicated, since the eigenstates have to be normalized before
acting with $\ln(v_1)$ and the normalization is done in $b_1$ space.
We hence proceed as
follows. After transforming the eigenstates to $b_1$ space, as in
eq.~\eqref{eq:etilde}, and doing the normalization in that space, as in eq.~\eqref{eprime},
 we transform back to $v_1$ space by means of an inverse Fourier
transform.
Since we are then back in $v_1$ space, we can act with $\ln(v_1)$ by multiplication, 
\begin{equation}
e^{\ln}_{\omega_1}(v_1):=\ln(v_1)\left[\mathcal{F}^{-1} {\tilde
    e}^\prime_{\omega_1} \right] (v_1).
\end{equation}
Finally, we apply again the
transformation in eq.~\eqref{eq:etilde} to go once more to $b_1$ space and
obtain
\begin{equation}\label{lnv1}
\widehat{ \ln(v_1)_{b_1}}{ {\chi}_{b_1}}(v_2,v_3)=\int {\rm d}\omega_2 {\rm d}\omega_3 \tilde{ {\chi}}_{b_1}^{\ln}(\omega_2,\omega_3)  e_{\omega_3}(v_3),
\end{equation}
where
\begin{equation}
\tilde{ {\chi}}_{b_1}^{\ln}(\omega_2,\omega_3):=\tilde{
  \Phi}(\omega_2,\omega_3) [\mathcal{F}  e^{\ln}_{\omega_1} ](b_1).
\end{equation}
Note that here we departed from the  procedure followed
in Ref.\cite{b1-madrid1}, where one acts with $\ln(v_1)$ before normalizing. The
errors introduced with that method can be neglected for semiclassical states
packed at large $\omega_i$. However, in this work we are interested in studying
more general cases. In the following analysis, we set $\tilde{  \Phi}(\omega_2,\omega_3)$ as a Gaussian
distribution centered at $\omega_2=\omega_2^*$ and $\omega_3=\omega_3^*$:
\begin{equation}
\tilde{
  \Phi}(\omega_2,\omega_3)=
\frac{1}{\sqrt{\pi}\sigma_2}
e^{-\frac{(\omega_2-\omega_2^*)^2}{2\sigma_2^2}} e^{i\beta_2\omega_2} \,
\frac{1}{\sqrt{\pi}\sigma_3}
e^{-\frac{(\omega_3-\omega_3^*)^2}{2\sigma_3^2}} e^{i\beta_3\omega_3} ~.
\end{equation}
Note that in principle we can choose more general states, such as squeezed and non-Gaussian states instead of above states. However, Gaussian states are desirable phenomenologically as  
they are peaked symmetrically in both of the conjugate phase space variables. In particular, the initial Gaussian states considered in our analysis will be such that they are peaked on classical trajectories 
specified by the values of classical phase space variables at initial `time' $b_i$ in a large macroscopic vacuum Bianchi-I spacetime. The above choice of Gaussian states thus allows us to understand the evolution to the 
quantum regime starting from the classical Bianchi-I spacetime. As the previous works on isotropic models in LQC have demonstrated, squeezed and more general states result in added features in the bounce regime 
due to the asymmetric quantum fluctuations of the states \cite{numlsu-3}. Similar results are being found for squeezed and non-Gaussian states for the loop quantization of vacuum Bianchi-I model in an independent study \cite{next}. 
It is to be noted that the main results of bounce and the physics of the Planck regime for the latter states share the same features as reported in this manuscript.

The above construction as presented in Ref. \cite{b1-madrid1}, provides a rigorous framework in the quantum theory to 
understand the detailed aspects of evolution in the loop quantized Bianchi-I spacetime. However, this involves various computational challenges. Before we proceed to understand and address them, let us note 
that there exists an effective spacetime description of the quantum theory in LQC. This description results in an effective Hamiltonian which has been derived rigorously in isotropic models for states 
which are sharply peaked at large volumes \cite{vt,psvt}. Following the latter results, the effective Hamiltonian constraint corresponding to eq.(\ref{C-q}) can be written as
\be\label{effham}
C_{\mathrm{eff}} = - 72 \pi \lp^4 \left(\sin(b_1) v_1 \sin(b_2) v_2 + \sin(b_2) v_2 \sin(b_3) v_3 + \sin(b_3) v_3 \sin(b_1) v_1 \right) ~ \approx ~ 0 ~.
\ee
Using Hamilton's equations modified dynamics in an effective continuum spacetime can be derived, resulting in 
\be\label{dotv1_v1}
\frac{\dot v_1}{v_1} = \frac{9 \pi \lp^2}{V} \, \cos(b_1) \left(\sin(b_2) v_2 + \sin(b_3) v_3 \right) ~,
\ee
and similarly for $v_2$, $v_3$ and $b_i$'s.

\section{Computational aspects}
In this section, we present various details of the computational aspects of our analysis. We start with the computational cost estimate 
to perform simulations for a wide variety of Gaussian states. To overcome the associated computational constraints led us to a 
\code{Cactus} \cite{Allen99a,cactus_web} implementation in our analysis. After explaining this implementation in Sec. IIIB, we summarize the scaling performance of our code 
on high performance computers used in our simulations in Sec. IIIC.

\subsection{Computational cost estimation}
In order to obtain the physical solution of the quantum Hamiltonian 
as described above, we first consider a Gaussian peaked at large eigenvalues $(\omega^*_2,\,\omega^*_3)$. At late times the expectation value of 
the physical observables would be such that the LQC trajectory agrees with 
the corresponding Wheeler-DeWitt trajectory. The physical wavefunction at any 
time $b_1$ is given by \eref{eq:chilqc}. Note that, the physical state is a 
3 dimensional object. However, in the numerical implementation, the state is
only stored at a single value of $b_1$ at a time as a two dimensional array of
size $n_2\times n_3$, where $n_2$ and $n_3$ respectively 
are the size of the spatial grid in the $v_2$ and $v_3$ directions.
In addition the evaluation of the integral in~\eref{eq:chilqc} we also
need to evaluate additional integrals in order to be able to calculate various
analysis quantities, such as eqs.(\ref{lnv2}) and (\ref{lnv1}). Denoting the total number of time-steps at which
we need to evaluate the state by $n$, the total number of times these
integrals need to be evaluated is $n\times n_2\times n_3$.

We numerically compute the integrals using a Gauss-Legendre integration, generally 
within the range of $\omega_a^*\pm10\sigma_a$. Here the integral is
approximated by a discrete sum with certain weights over frequency values
within the given range of integration. Let us say we choose $n_{\omega_a}$
frequency values in each $\omega_a$ direction. Then to evaluate the integral 
at one spatial grid point we need to sum over 
$n_{\omega_2}\times n_{\omega_3}$ terms. As described in
section~\ref{sec:cactus} adding the contribution to all the integrals at each
frequency point requires $44$ floating point operations.

Therefore, the total number of floating point operations in the computation of 
the integrals for $n$ values of $b_1$ and all grid points  $v_2$ and $v_3$ 
will be equal to
\be
N = 44 \times n \times n_2\times n_3\times n_{\omega_2}\times n_{\omega_3}.
\ee
Let us now consider a typical example of a simulation of sharply peaked 
Gaussian state with $\omega_2=1000$, $\omega_2=1000$, $\sigma_2=50$, 
$\sigma_3=50$. This would typically require: $n=1024=2^{10}$, 
$n_2=n_3=4096=2^{12}$ and $n_{\omega_2}=n_{\omega_3}=256=2^8$. 
Then the total number of floating point operations would be 
\be
N = 44\times 2^{50} \approx 5\times 10^{16} {\rm flops}.
\ee
On a modern workstation with a peak performance of $50$ GFlops per second, the
total computation time in seconds would be 
\be
T_{\rm comp} \approx \f{N}{50\times 10^9} \approx 10^6~ {\rm sec} \approx 11.5 ~{\rm days}.
\ee
On the other hand the simulation of a widely spread state would 
require larger grids in $b_1$, $v_2$ and $v_3$ which would significantly 
increase the computation time proportionally. Clearly such simulations are not
suited for a single workstation. In addition to the long computation time,
there should be enough memory available at the workstation to store the
necessary eigenfunction data. For the grid size considered above one 
would need approximately $0.5$ TB of random access memory. 
This memory requirement is clearly beyond a typical modern workstation and the
need for a parallel implementation suitable for high performance computing
platforms is clear. In the following section~\ref{sec:cactus} we will describe
in more detail our parallel implementation.

\begin{figure}[t!]
  \begin{minipage}[b]{0.49\textwidth}
    \includegraphics[angle=0,width=0.99\textwidth,height=!,clip]{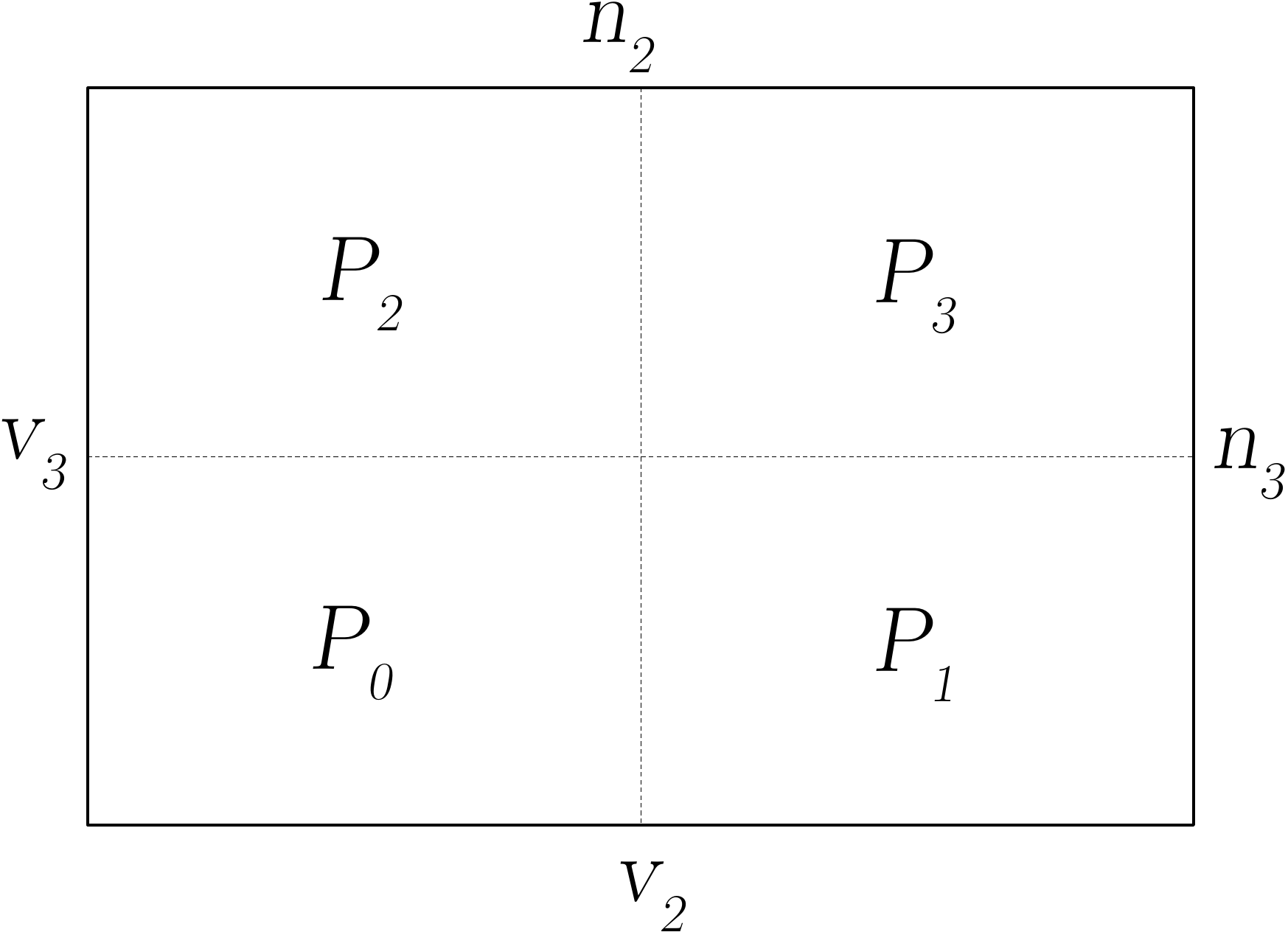}
  \end{minipage}
  \begin{minipage}[t]{0.49\textwidth}
  \includegraphics[angle=0,width=0.99\textwidth,height=!,clip]{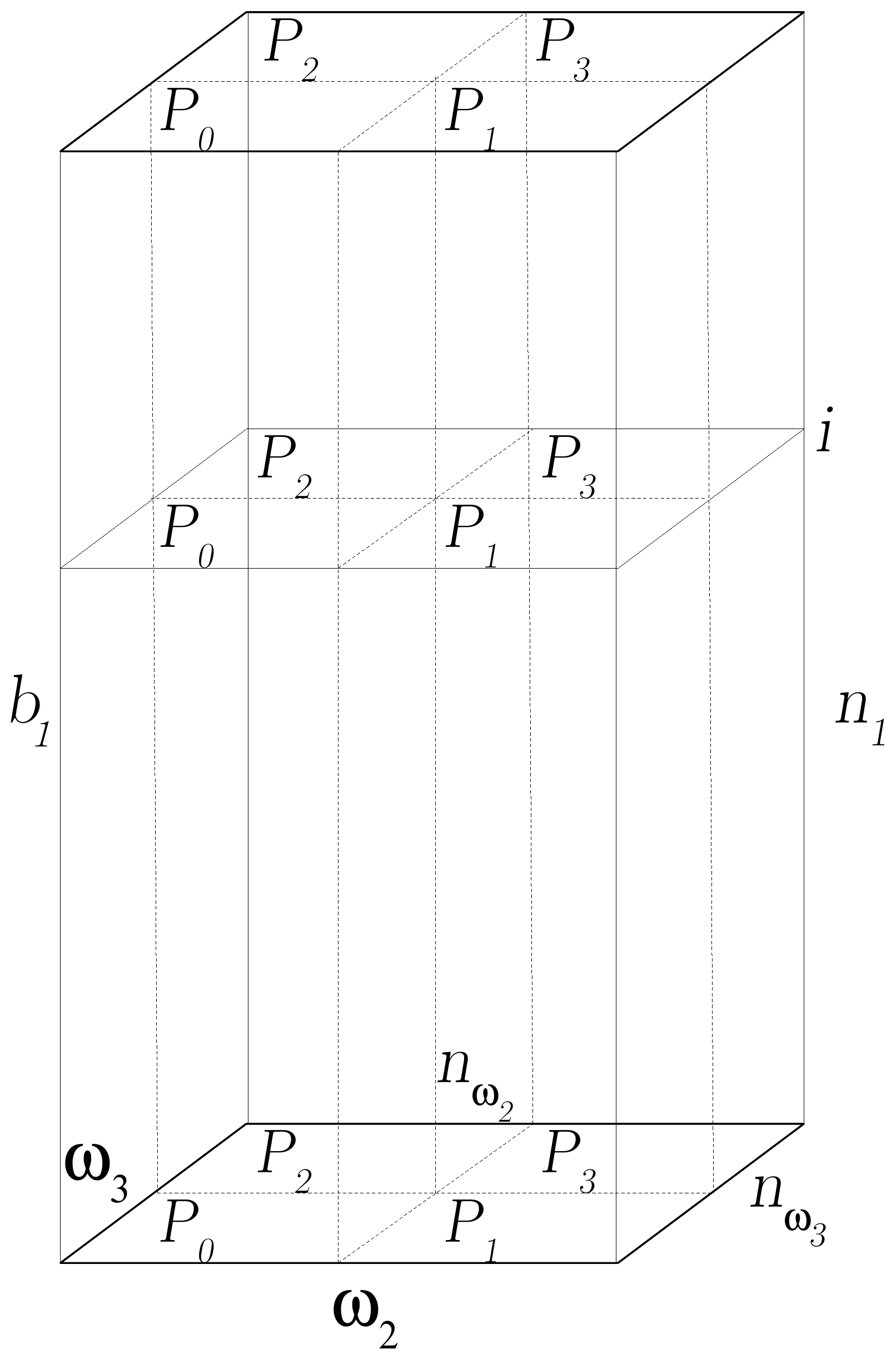}
  \end{minipage}
  \caption{Illustration of the processor distribution of $\chi_{b_1}(v_2,v_3)$
	   (left) and of $\tilde{\chi}_{b_1}(\omega_2,\omega_3)$ (right) when
	   using 4 MPI processes. In the left plot, $v_2$ and $v_3$ labels
	   the direction in volume space while $n_2$ and $n_3$ denotes
	   the number of grid points in those directions.
	   Similarly, in the right plot $\omega_2, \omega_3$ labels the
	   frequency space directions and $b_1$ labels the time direction, while
	   $n_{\omega_2}, n_{\omega_3}$ and $n_1$ denotes the number of
	   grid points. A slice through the data structure corresponding to
	   a single representative ${b_1}_i$ value is illustrated with the
	   horizontal plane labeled $i$.
  }
  \label{fig:parallel}
\end{figure}

\subsection{Cactus implementation}
\label{sec:cactus}

\begin{figure}
  \begin{minipage}[b]{0.555\textwidth}
    \includegraphics[angle=0,width=0.99\textwidth,height=!,clip]{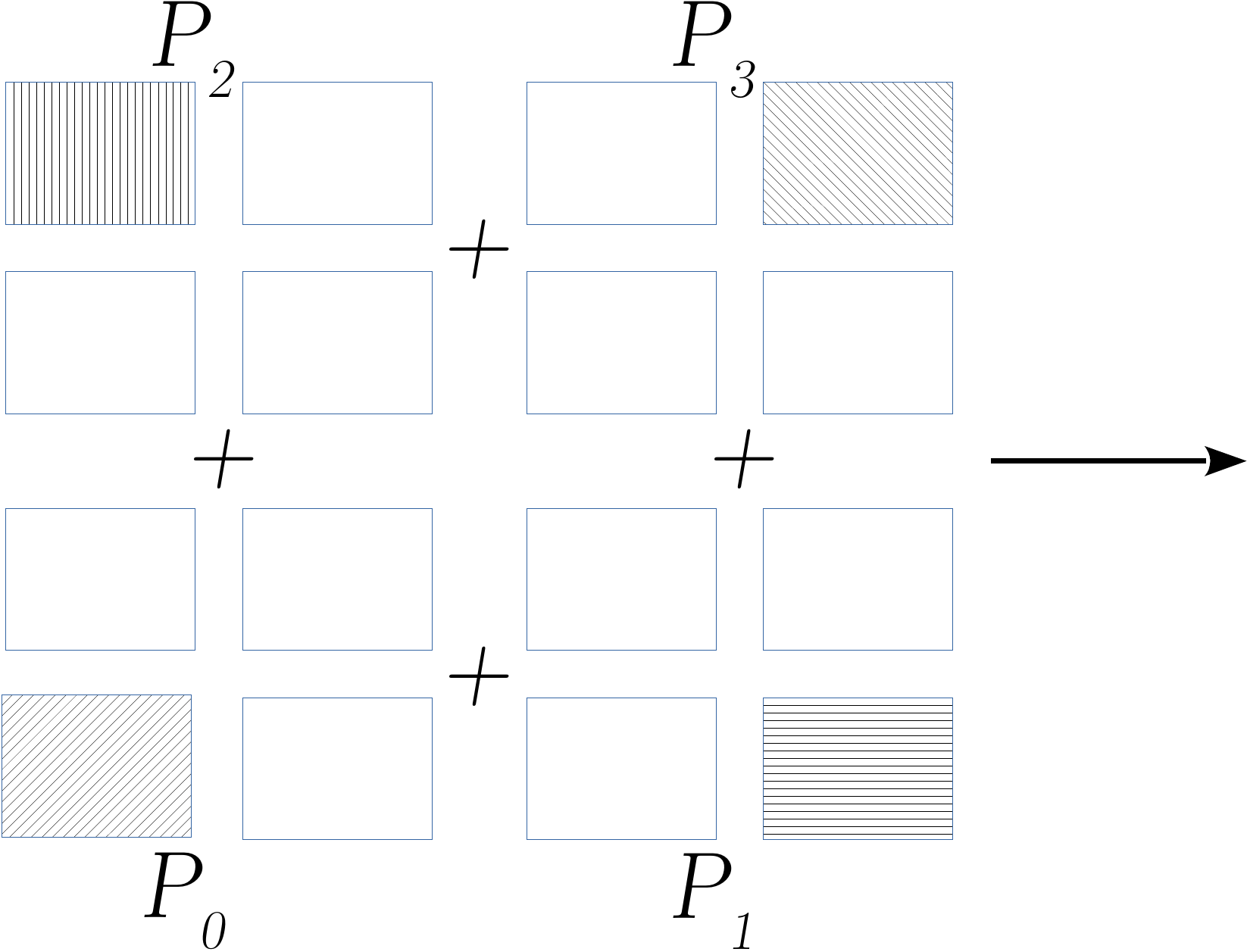}
  \end{minipage}
  \begin{minipage}[b]{0.425\textwidth}
    \includegraphics[angle=0,width=0.99\textwidth,height=!,clip]{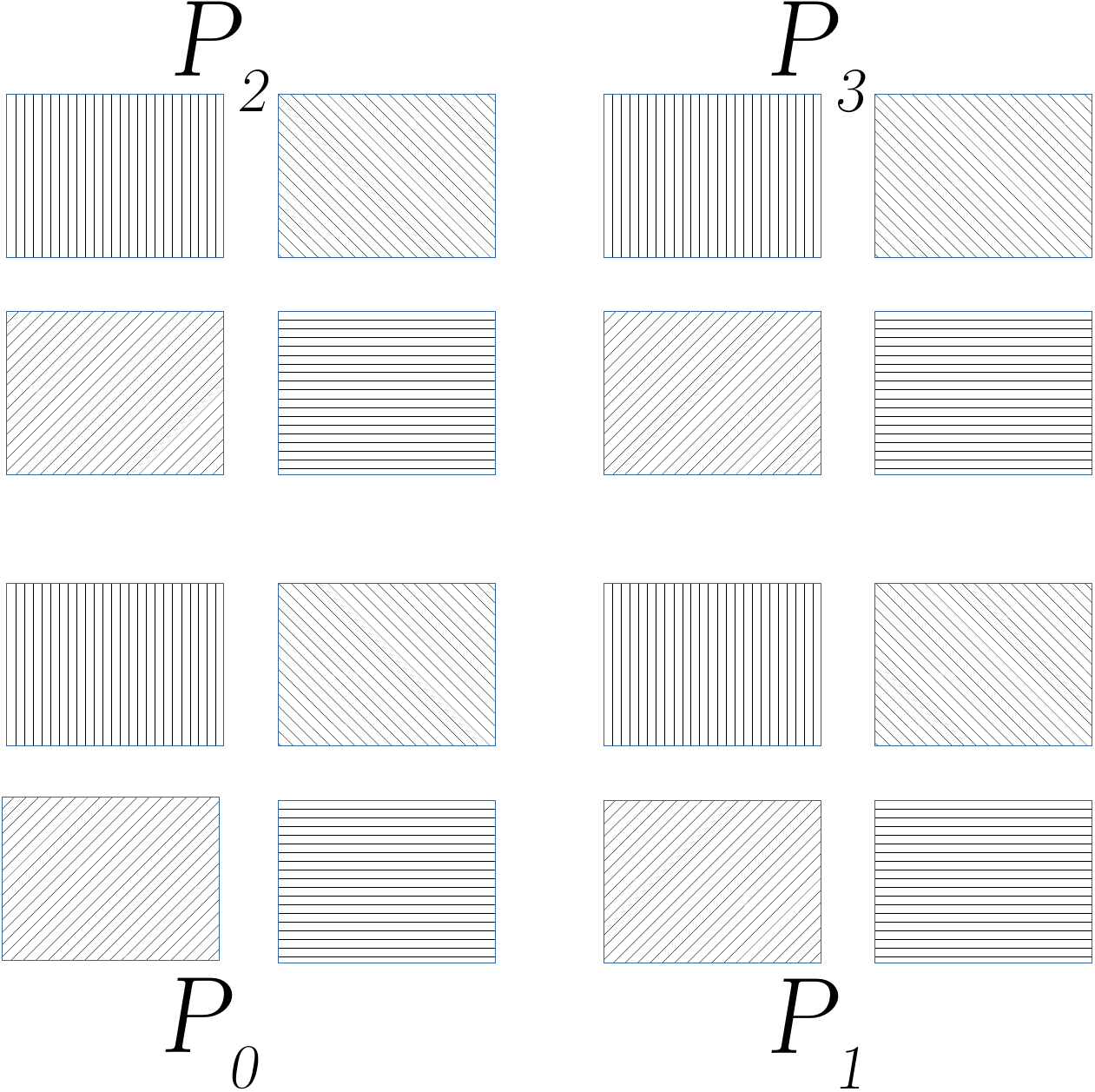}
  \end{minipage}
  \caption{Plot showing  sum reduction used to communicate all frequency
	   information to all MPI processes (for the case of 4 MPI processes).
	   Before the reduction (left side) each MPI process only knows a
	   part of the data (indicated by the different hatched regions) and
	   copies that to the correct location in a temporary array of full size
	   ($n_{\omega_2}\times n_{\omega_3}$) with zeros everywhere else
	   (indicated by blank regions). After the sum reduction (right side)
           all MPI processes has identical frequency information.}
   \label{fig:reduction}
\end{figure}

We have implemented the numerical evaluation of the state~\eref{eq:chilqc}
and the additional integrals needed for calculating analysis quantities as
a thorn in \code{Cactus} \cite{Allen99a,cactus_web}. At any given time $b_1$, the state 
$\chi_{b_1}(v_2,v_3)$  can be stored as a 2-dimensional grid function of size 
$n_2\times n_3$ (see left plot in Fig.~\ref{fig:parallel}). The computation
of $\chi_{b_1}(v_2,v_3)$ is therefore parallelized in the $v_2$ and $v_3$
directions. On the other hand the discrete representation
of $e_{\omega_2}, e_{\omega_3}$ and $\tilde{\chi}_{b_1}$ will typically have
different sizes and can therefore not be stored as grid functions (grid
functions in \code{Cactus} all have the same size). In particular
$\tilde{\chi}_{b_1}(\omega_2,\omega_3)$ of size
$n_{\omega_2}\times n_{\omega_3}\times n_1$ has to be calculated
initially (using numerical fast Fourier transforms (FFTs)) and stored for the remainder of the
simulation (see right plot in Figure~\ref{fig:parallel}).  This is the largest
object considered by far.  For $n_{\omega_2}=n_{\omega_3}=256$ and $n_1 = 131072$, this
single object requires 128 GBytes of storage (this array is double precision
complex). We store this object as a vector of grid arrays that is distributed
among processors in the $\omega_2$ and $\omega_3$ direction but not in the $b_1$
direction. That is, each processor owns a range of $\omega_2$ and $\omega_3$
values and for each pair of frequencies has data for all $b_1$ values. This is
illustrated for the case of 4 message passing interface (MPI) processes in the right plot in 
Fig.~\ref{fig:parallel}. This allows us to calculate the FFT's in parallel
by calling serial FFT routines for different $(\omega_2, \omega_3)$ values 
on different processors. At
the time when the integral has to be evaluated, each processor (computing a
chunk of the $v_2$ and $v_3$ grid) needs all frequency values (but only for
that single $b_1={b_1}_i$ value). Therefore, before each integral evaluation, we
need to communicate the ${b_1}_i$ slice of $\tilde{\chi}_{b_1}(\omega_2,\omega_3)$ to
all processors.  This operation is performed using a special \code{Cactus} local array
sum reduction. With each MPI process owning a local array (the size has to be
the same on all MPI processes) this sum reduction will, for each element of the
array, add up the contribution from all MPI processes and place the resulting
array on either a single MPI process or alternatively on all MPI processes.
Thus, on each MPI process, we allocate a local 2-dimensional array of size 
$n_{\omega_2}\times n_{\omega_3}$ and initialize it to zero everywhere except
for the chunk of frequency values where the MPI process knows the value of
$\tilde{\chi}_{b_1}(\omega_2,\omega_3)$. With a call to the \code{Cactus} local array
sum reduction we finally make sure that the result of the sum is communicated
to all MPI processes. This is illustrated in Fig.~\ref{fig:reduction}.

The numerical evaluation of the integral~\eref{eq:chilqc} then consists
of looping over all grid points (in $v_2$ and $v_3$) and for each grid point
we have a double loop over the frequency directions ($\omega_2$ and
$\omega_3$). In this loop we need 8 floating point operations to evaluate and
add the contribution from this frequency pair to the integral~\eref{eq:chilqc}.
For analysis purposes we need to evaluate 2 more integrals of this cost and
5 more at half the cost. The half cost integrals are due to the fact that
their integrands are given by a real number multiplied one of the previously
calculated integrands, leading to only 4 floating point operations. Thus
each loop consists of 44 floating point operations with a perfect mix of
22 multiplications and 22 additions.

After the state has been calculated, we then have to calculate various 
expectation values. These are given by discrete sums over the state itself
as well as the other gridfunctions calculated in the integration loop. For 
these we perform local sums over the patch of the grid owned by each MPI
process and again resort to using the \code{Cactus} local array sum reduction. In
order to reduce the communication overhead and latency, we perform the
reduction of these analysis quantities at the same time as we prepare the
integrands for the next time step as described earlier.

In order to be able to use modern high performance computing (HPC) architectures, we have ported the
integration routine to work on both graphical processing units (GPUs) and
Intel Xeon Phis (codename Knights Corner). In the first case we use the open accelerators (OpenACC) 
programming paradigm and in the second case open multi-processing (OpenMP) with Intel's offload 
compiler directives. These programming paradigms are very similar to each 
other in the sense that you use compiler directives to specify when and
what data to transfer to and from the device and to indicate which loop
to run in parallel. In both cases we are also able to use the CPUs on the
system at the same time. We accomplish this, by splitting the loop into
two non-overlapping pieces. One piece is run on the CPU (using OpenMP
parallelization to ensure that all CPUs are used) while the other
is run on the accelerator (either GPU or Xeon Phi) simultaneously. Since
it is not a priori known what the performance of the accelerator is compared
to the CPU we have implemented an automatic work distribution scheme in the
\code{Cactus} code. We do this by adjusting, at runtime, the work distribution
between the CPU and the accelerator
using the number of outer loop iterations to run on the CPU as an
optimization parameter. In order to minimize the run time we first need to
bracket the minimum of runtime as function of the amount of CPU work. We do
this by performing the integration at the first iteration completely
on the accelerator (leaving the CPU idle) and measure the time. On the next
iteration we do all the work on the CPU (leaving the accelerator idle) and
again measure the time. We then do half the work on the CPU and half the work
on the accelerator (concurrently) and measure the time. If we have not yet
bracketed the minimum in time, we adjust the workload appropriately until we do and
then continue with a golden section search~\cite{kiefer53a} for the minimum.
This certainly
expends a few computational resources initially, but guarantees that we will
use all computational resources after a few (order 10) iterations.

The total number of floating point operations in the main
computational kernel can be estimated to be
\[
  N_{\mathrm{fp}} = 44 \times n \times n_2\times n_3\times n_{\omega_2}
  \times n_{\omega_3},
\]
where $n_2$ and $n_3$ are the number of grid points in the $v_2$ and
$v_3$ directions and $n_{\omega_2}$ and $n_{\omega_3}$ are the number of
integration points in the frequency domain, $n$ is the number of time steps
where we actually evaluate the state. The number 44 comes from the number
of floating point operations needed to update all the integrals in the
innermost loop.  We have also
used PAPI (performance accessing hardware interface) for accessing hardware counters to measure the actual number of
floating operations generated by the compiler and find good agreement between
the measured values and the estimated values, indicating that the compiler
does not generate superfluous floating point instructions. 

\begin{figure}
  \begin{minipage}[b]{0.49\textwidth}
    \includegraphics[angle=0,width=0.99\textwidth,height=!,clip]{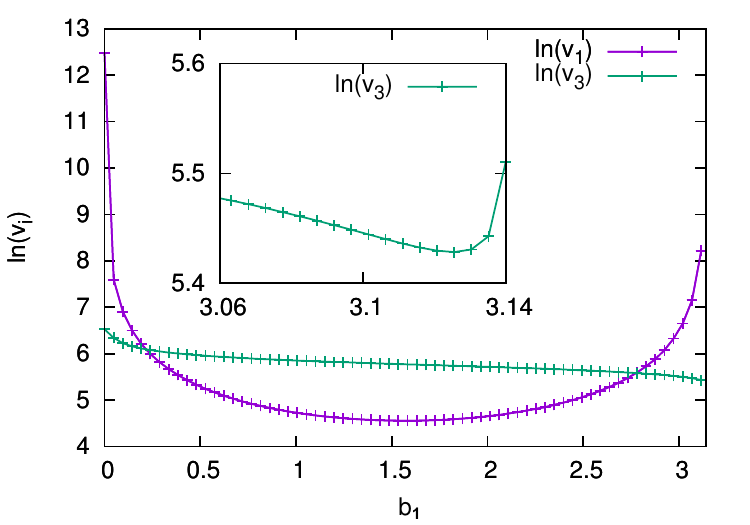}
  \end{minipage}
  \begin{minipage}[b]{0.49\textwidth}
    \includegraphics[angle=0,width=0.99\textwidth,height=!,clip]{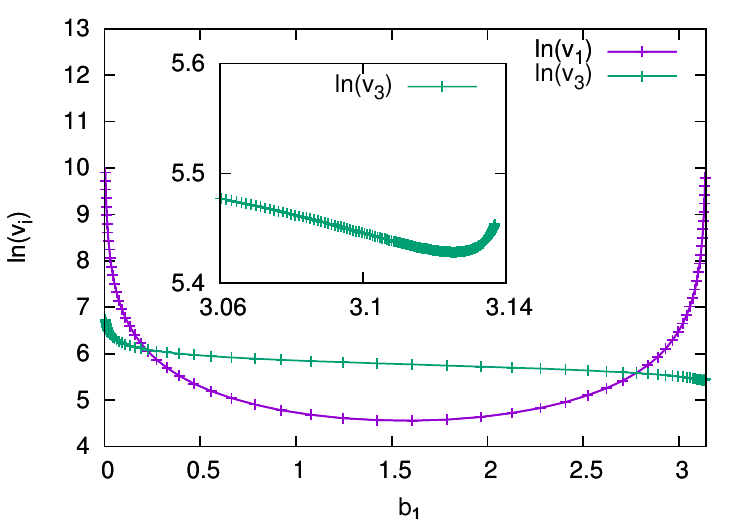}
  \end{minipage}
  \caption{Both left and right plots show the expectation values of $\widehat{\ln(v_1)|_{b_1}}$
	  (purple (dark) curve) and $\widehat{\ln(v_3)|_{b_1}}$ (green (light) curve) as functions of $b_1$ for a
	   case where we only evaluate the state at 1 in every 200 possible
	   values of $b_1$ and the bounce in $v_3$ occurs close to $b=\pi$. In
	   both, the main plot only show 1 in every 10 of the $b_1$ values where
	   the state is evaluated, while the inset (showing a zoom of the
	   bounce in $v_3$) includes all points.}
   \label{fig:arc}
\end{figure}

In principle we could evaluate the state at all $n_1$
possible $b_1$ values, however this is computationally expensive and not really
necessary as long as we have sufficient evaluations to resolve the minima in
$\ln(v_2)$ and $\ln(v_3)$. Thus we typically use $n\ll n_1$ and therefore
have some freedom in choosing at which of the evenly spaced
values of $b_1$ ($\Delta b_1=\pi/n_1$) we want to evaluate the state. The
easiest choice would be to do the evaluation for every $m=n_1/n$ values of $b_1$
(i.e.\ uniform distribution in $b_1$). However, sometimes the  bounce
in either $v_2$ or $v_3$ happens over a small range of $b_1$-values near
$b_1=\pi$. This is illustrated in the left plot of Figure~\ref{fig:arc},
which shows the expectation values of $\widehat{\ln(v_1)|_{b_1}}$ (purple (dark) curve) and $\widehat{\ln(v_3)|_{b_1}}$
(green (lighter) curve) as functions of $b_1$ for a simulation with $m=n_1/n=200$. In the
main plot, we only show every 10 values of $b$ for which the state was
evaluated, while in the inset we show all values. The inset is a zoom in
of the region of $b_1$ near $\pi$ where the bounce in $v_3$ happens. As can
be seen, the bounce in $v_3$ is not very well resolved when uniform sampling
is used.

We know that the expectation values of all $v_i$ diverge near both 0 and
$\pi$. Therefore, if we use a constant spacing in arc-length along the 
$\ln(v_1)$ versus $b_1$ curve we will obtain a non-uniform spacing of $b_1$ values
with a significantly higher density of points near 0 and $\pi$ than at the
midpoint of the interval. It turns out
that the dependence of $\ln(v_1)$ as function of $b_1$ can in general be
approximated by the function
\[
g(b_1) = \ln\left (\frac{1}{\sin b_1}\right ) + C,
\]
where constant $C$ depends on the physical parameters. As expected, this function has a
minimum (corresponding to the bounce in $v_1$) at $b_1=\pi/2$ but diverges for
$b_1=0$ and $b_1=\pi$. The arc-length distance $S({b_1}_{(0)},{b_1}_{(f)})$ along the curve $g(b_1)$ between ${b_1}_{(0)}$ and ${b_1}_{(f)}$
is
\[
  S({b_1}_{(0)},{b_1}_{(f)}) = \int_{{b_1}_{(0)}}^{{b_1}_{(f)}} \sqrt{1+\left (\frac{d g}{db_1'}\right )^2}\, db_1' =
  \int_{{b_1}_{(0)}}^{{b_1}_{(f)}} \sqrt{1+\frac{\cos^2 b_1'}{\sin^2 b_1'}}\, db_1' =
  \int_{{b_1}_{(0)}}^{{b_1}_{(f)}} \frac{1}{\sin b_1'}\, db_1' = \ln\left (
\frac{\tan\left (\frac{{b_1}_{(f)}}{2}\right )}{\tan\left (\frac{{b_1}_{(0)}}{2}\right )}\right ).
\]
This function can be inverted
\[
  b_1({b_1}_{(0)},S) = 2\tan^{-1}\left [e^S \tan\left (\frac{{b_1}_{(0)}}{2}\right )\right ].
\]
to give the value of $b_1$ for a point a given arc-length distance, $S$, away
from the starting point ${b_1}_{(0)}$. We choose our starting point as ${b_1}_{(0)}=\pi/n$ 
and then calculate the arc-length distance between this point and the mid
point, $S_{\mathrm{mid}}=S(\pi/n,\pi/2)$, of the curve. To setup the
evaluation points for the first half interval, we set $\Delta S = 
2 S_{\mathrm{mid}}/(n-1)$ and, for $i=0,n/2$, we define a list of desired
evaluation points ${b_1}_i = b_1({b_1}_{(0)},(i-1)\Delta S)$. From the points in our equally
spaced $b_1$-values we then select the $n/2+1$ points that are closest to the
desired evaluation points. The remaining points are then found by symmetry.
Repeating the simulation shown in the left plot of Fig.~\ref{fig:arc}
with this non-uniform choice of $b_1$ values is shown in the
right plot (produced in exactly the same way as the left plot) of
Fig.~\ref{fig:arc}. We find that the the bounce in $v_3$ is very well resolved.

\begin{figure}[t!]
  \includegraphics[width=0.49\textwidth]{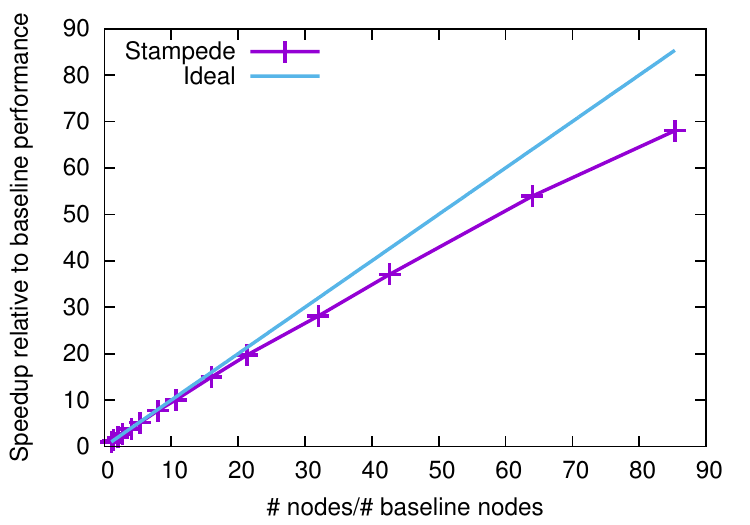}
  \includegraphics[width=0.49\textwidth]{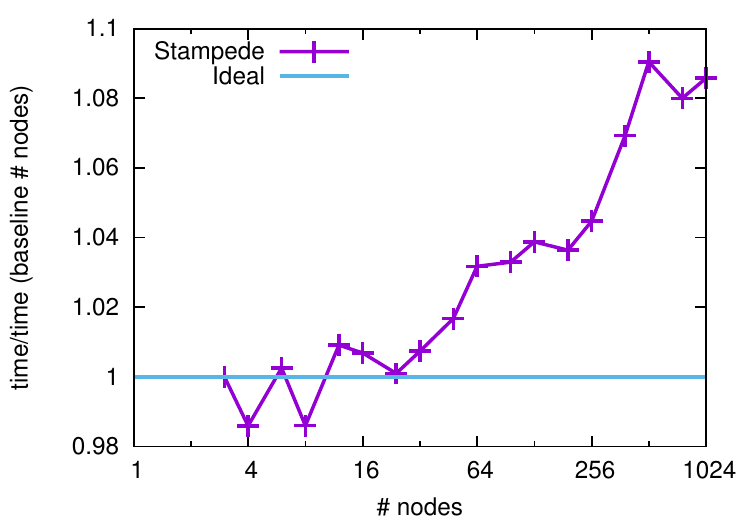}
  \caption{Strong (left plot) and weak (right plot) scaling for the MPI 
           parallelized \code{Cactus} code based on timings on the XSEDE resource
           Stampede. Strong scaling is shown as the time on 3 nodes divided by
           the time on $x$ nodes (i.e.\ speedup relative to the performance on 3
           nodes) for the same amount of total work. Here the ideal scaling is
           indicated by the line with slope 1 (doubling the number of nodes
           should halve the time). Weak scaling is shown as 
           the time on $x$ nodes divided by the time on 3 nodes when the total
           amount of work per node is kept constant. Here ideal scaling is
           indicated by the horizontal line with value 1 (it should take the
           same amount of time to do twice the amount of work on twice the 
           number of processors).}
  \label{fig:scaling}
\end{figure}

\subsection{Performance and scaling}

The computational kernel is vectorized and performs very efficiently on
current CPU architectures. We have measured the performance and found that the
kernel performs at about 60\% of theoretical peak on a single core and at about
50\% of peak on a 16 core shared memory node using OpenMP parallelization
(there is probably a bit of memory contention as well as OpenMP parallelization
overhead). The performance on a Nvidia Tesla K20X GPU is about 25-30\% of peak
and on the Intel Xeon Phi about 15-20\%. The lower performance on the Xeon Phi
compared to a Intel CPU is caused by a larger number of Level 1 data cache
misses. This is due to the fact that on the Xeon Phi we get the best 
performance when running 2 threads or more per core (at least 2 is required in order
to keep the floating point units busy) and each physical core
has the same level 1 data cache as an Intel CPU core. Thus on the CPU, each
thread has sole access to the full 32 KB level 1 data cache, while on the Xeon
Phi multiple threads (in fact we obtain the best performance when using 4 threads
per core) share the 32 KB level 1 data cache. We have experimented with
different schemes for blocking for better cache uses, but have not yet found
any scheme that leads to improved performance.

In Fig.~\ref{fig:scaling} we show the strong and weak scaling as measured on
the XSEDE resource Stampede. Here we used the full nodes (16 CPU cores and 1
Intel Xeon Phi accelerator card).

For the strong scaling results we used 
representative values $n_2 = n_3 = 8192$ and $n_{\omega_2} =
n_{\omega_3} = 256$. However, $n_1=16384$ was chosen to be much smaller than
the value that would be used in production runs, in order for the job to fit
in memory on 3 nodes. The strong scaling plot (left plot) obtained at the XSEDE
resource Stampede shows the speedup
relative to the performance on 3 nodes as a function of the number of nodes
used divided by 3.  The timings are based on the total wall time
as reported by \code{Cactus} timers (including startup, initialization and evolution).
As the total amount of work is kept constant, ideal scaling
is a straight line with slope 1. As we increase the number of nodes from 3 to
256 (from 48 cores and 3 Xeon Phis to 4096 cores and 256 Xeon Phis) the
code runs 68 times faster. Ideal scaling would result in a speed up of 85.33.
This is very good strong scaling.

For the weak scaling results, the parameters for the 3 node baseline run were
$n_2 = 1024$, $n_3 = 1536$, $n_{\omega_2} = n_{\omega_3} = 256$ and
$n_1=16384$ (again $n$ was chosen so that the job would fit on 3 nodes). As
the number of nodes was increased the grid size ($n_2$ and $n_3$)
was increased proportionally with all other parameters kept the same. Thus
in each case, every node got assigned a piece of the grid of size 
$1024\times 512$ and at 1024 nodes we used $n_2 = 16384$ and 
$n_3 = 32768$. In this case ideal scaling should result in constant
runtime, independent of the number of nodes, i.e.\ a horizontal line with
value 1. As can be seen from the right plot in Fig.~\ref{fig:scaling}, when
increasing the node count from 3 to 1024 (a factor of 341.33) the code slows
down by less than 10\%. The largest weak scaling job used 16,384 CPUs and
1024 Xeon Phis. Users of Stampede can only get access to this queue (the normal
queue tops out at 256 nodes) after providing evidence of being able to run at
scale. Thus we are able to run with less than 10\% loss of efficiency when
running on the largest number of nodes available to jobs in a standard queue
(even larger jobs can run upon special request).

\section{Results}
Using the \code{Cactus} implementation discussed in Sec. III, more than a hundred simulations were performed for 
different initial conditions corresponding to various choices of $\omega_2^*$ and $\sigma_2$. Due to the richness of presence of 
various dimensions and parameters, a variety of parameters can be changed at once. We focused our analysis on keeping all but few parameters fixed. 
The reason for this is tied to the underlying symmetry in the Hamiltonian constraint, where three directions are on equivalent footing and only the difference of $\omega_i$'s matters.  In order to 
extract and understand various physical results in the following the value of $\omega_3^*$ was fixed to 
$\omega_3^* = 1000$ with $\sigma_3 = 40$. The value of $\omega_1$ was determined using the Hamiltonian constraint. 
Further, the phases $\beta_2$ and $\beta_3$ are fixed to 0.1.

\begin{figure}[t!]
 \includegraphics[angle=0,width=0.5\textwidth]{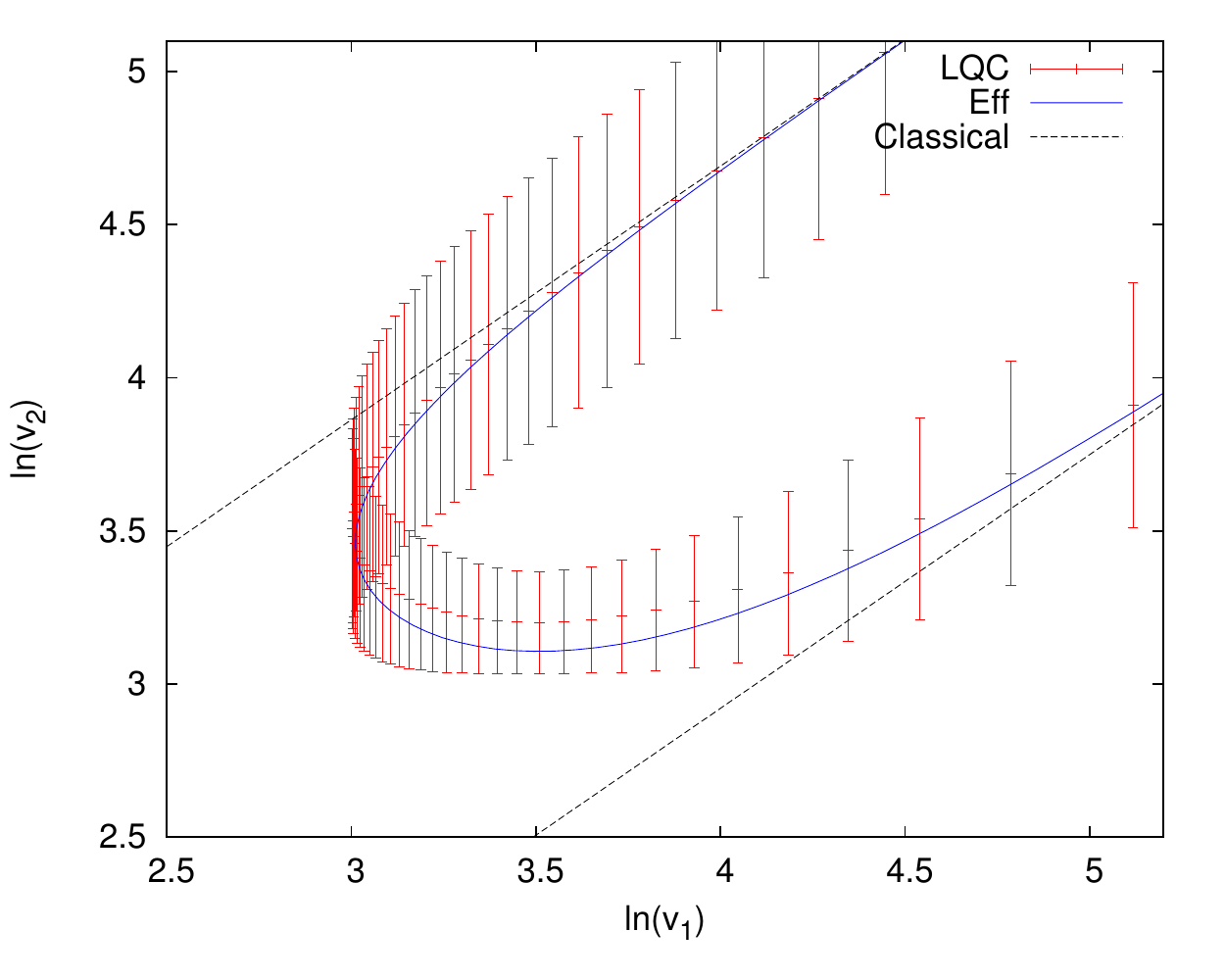}
 \caption{This plot shows a typical singularity resolution in the relational dynamics where the expectation values along with dispersion of $\widehat{\ln(v_2)|_{b_1}}$ 
are plotted versus those of $\widehat{\ln(v_1)|_{b_1}}$. The effective dynamics provides a good approximation throughout the evolution. For 
comparison, the singular and disjoint classical solutions are also shown to which the quantum expectation values asymptote at small 
spacetime curvature. The state parameters are $\omega_2^* = 100$ with $\sigma_2 = 14$, and $\omega_3^* = 1000$ with $\sigma_3 = 40$. }
 \label{fig:lnv2_lnv1}
\end{figure}

In the following we first demonstrate the resolution of singularity for some states, the agreement with the classical theory at large volumes and the way effective dynamics provides an excellent approximation for such states. This is followed by discussing the evidence of departure between the 
quantum theory and effective dynamics in Sec. IVB. These departures are studied and quantified using different parameters in Sec. IVC. In Sec. IVD, we apply the 
results of numerical computation of expectation values to estimate the expansion and shear scalars in this spacetime.

\begin{figure}[t!]
 \includegraphics[angle=0,width=0.48\textwidth]{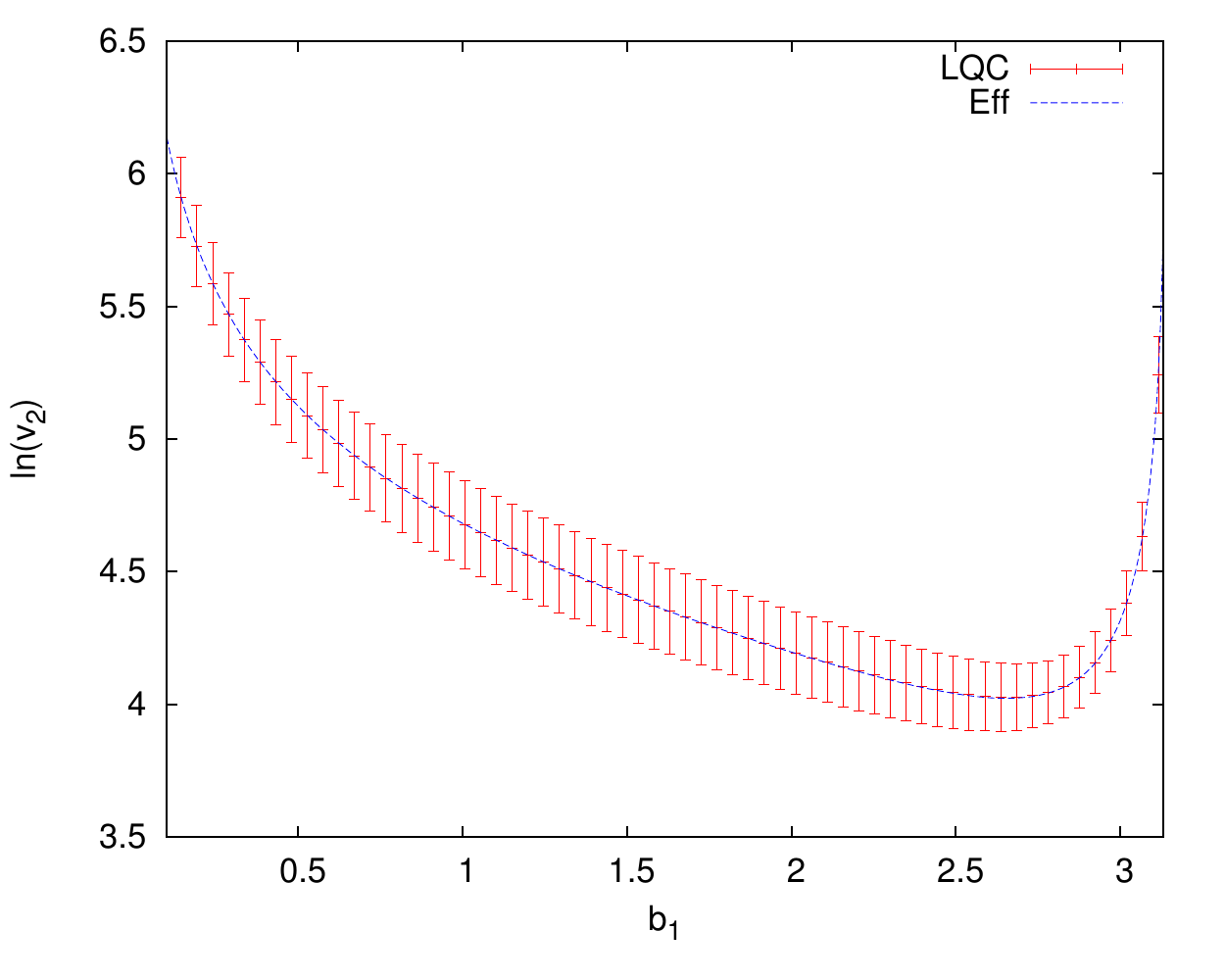}
 \includegraphics[angle=0,width=0.48\textwidth]{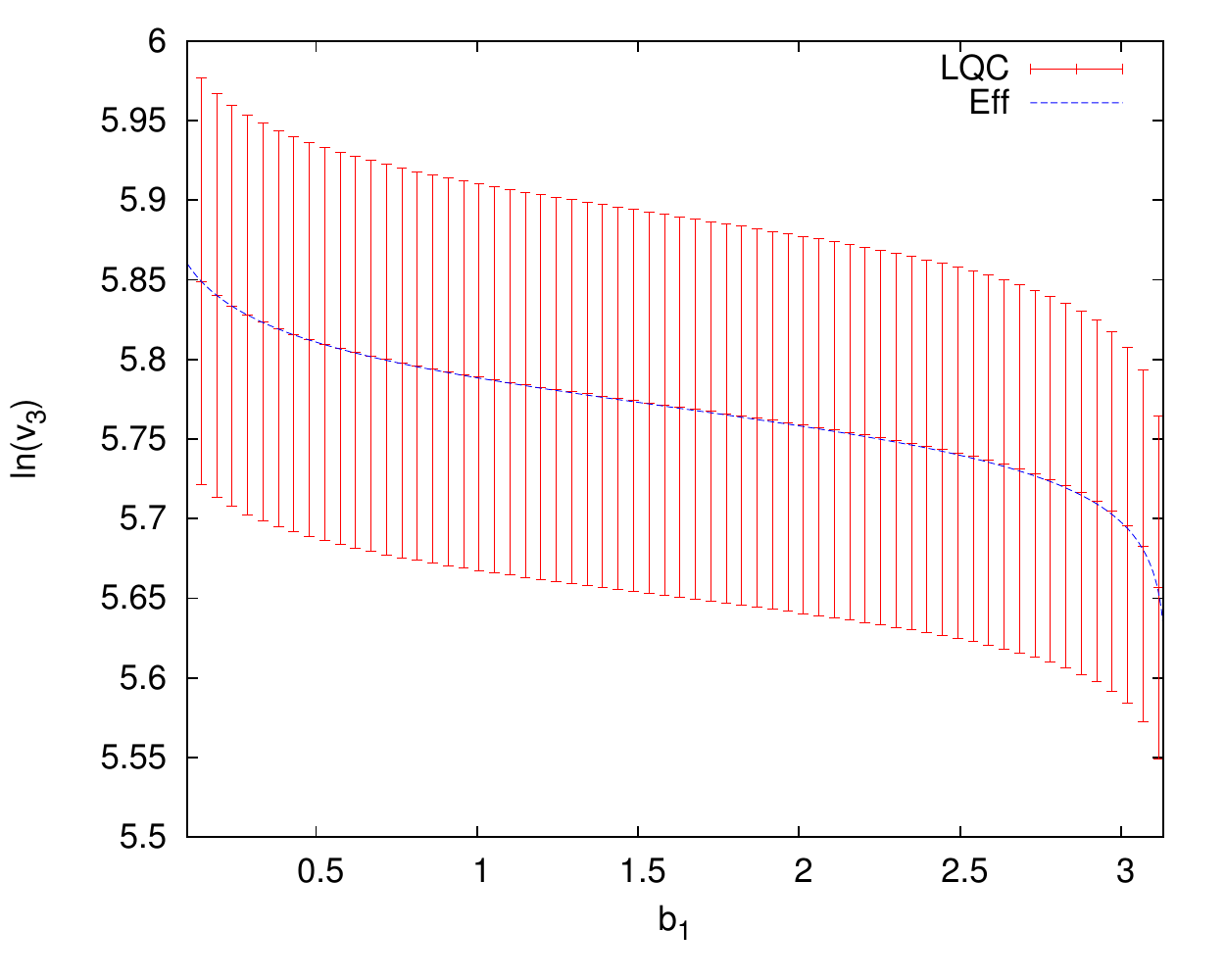}
 \caption{Evolution of expectation values of $\widehat{\ln(v_2)|_{b_1}}$ and $\widehat{\ln(v_3)|_{b_1}}$ versus $b_1$ is shown. Parameters are $\omega_2^* 
= 250$ with $\sigma_2 = 45$, and $\omega_3^* = 1000$ with $\sigma_3 = 40$. As can be seen, the effective dynamics is in excellent agreement 
with the quantum dynamics.}
 \label{fig:lnv2lnv3_b1_250}
\end{figure}

\subsection{Singularity resolution}
For all the simulations carried out in our analysis, classical singularity is found to be 
resolved in the quantum theory. The classical big bang singularity is replaced by non-singular evolution
of volumes $v_i$ in time $b_1$. The approach to singularity in the classical theory for the vacuum Bianchi-I 
spacetime is not point like as in the isotropic models, but is a cigar like. The initial conditions in the anisotropic evolution are 
such that two of the directional scale factors bounce and the third undergoes a recollapse.  Results from a representative simulation are shown in 
Fig. \ref{fig:lnv2_lnv1}, where we have plotted relational observables $\ln(v_2)$ versus $\ln(v_1)$ in the quantum theory, effective dynamics and classical theory. Let us first focus on the expectation values of  
quantum operators and compare them to the classical solutions. We find that starting from the upper classical branch, where initial conditions for the quantum 
evolution are given at the large volumes the classical and 
quantum curves agree for a certain period. As the classical curve approaches singularity, there is a departure between the classical and quantum theory. The classical curve continues evolution to the singularity whereas the quantum dynamics results in 
a non-singular turnaround. After the bounce, when the volume become large in the subsequent evolution the quantum curve again approximates a classical solution. The upper and lower classical solutions are disjoint and singular, which are bridged by the quantum theory.

\begin{figure}[b!]
 \includegraphics[angle=0,width=0.5\textwidth]{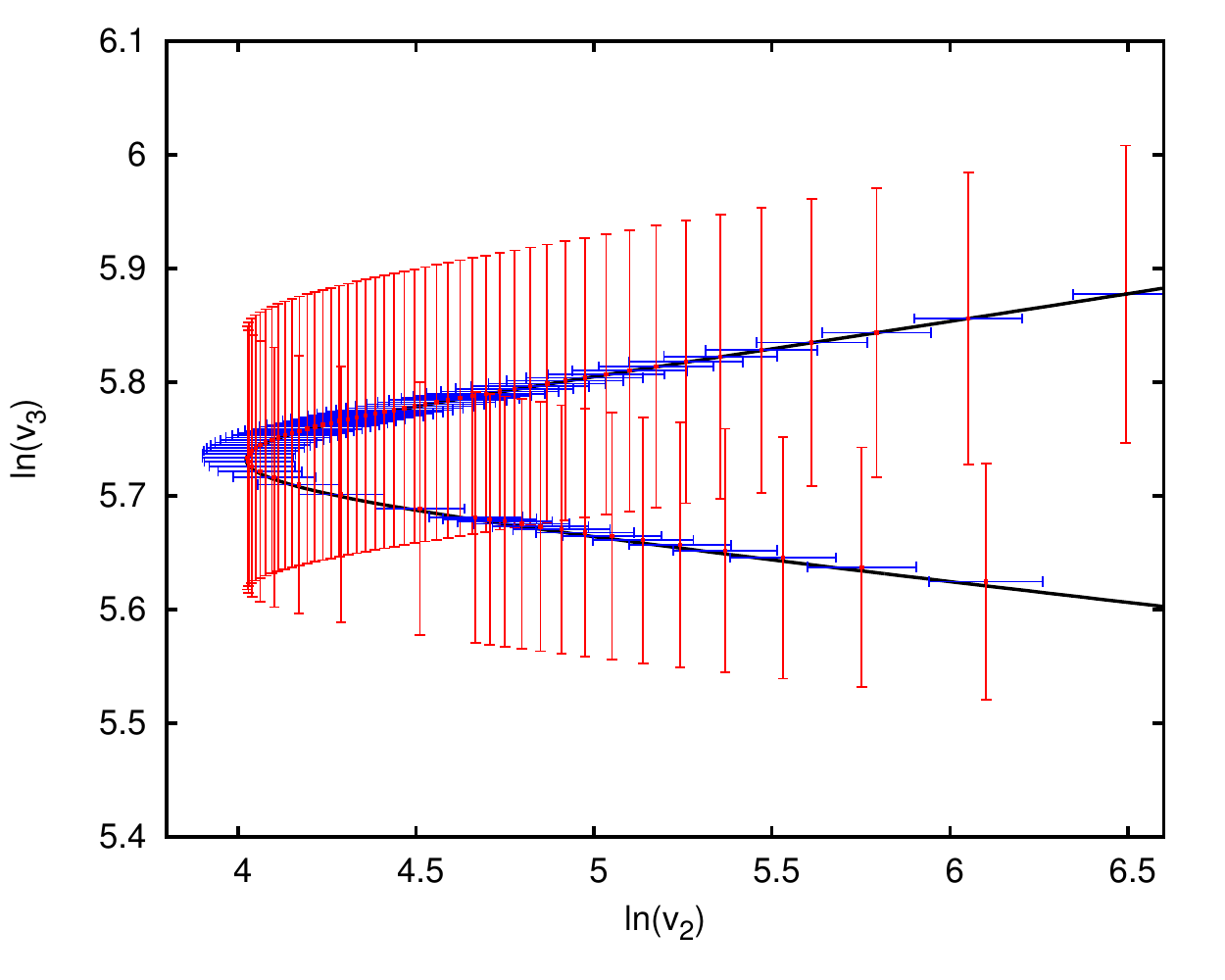}
\caption{This plot shows the relational observables corresponding to  $\ln(v_3)$ and $\ln(v_2)$ in the quantum theory and their behavior in effective dynamics (shown by the solid curve). Parameters are same as in Fig. \ref{fig:lnv2lnv3_b1_250}. Dispersions for both the observables are shown.}\label{fig:v3v2_250}
 \end{figure}

Now let us analyze the effective dynamics trajectory in relation to the quantum expectation values for the above simulation.  The trajectory obtained from the 
effective Hamiltonian constraint provides a very good approximation to the quantum dynamics. As one can see, the effective dynamical trajectory remains close to the mean value 
obtained from the quantum dynamics. (For larger values of $\omega_2^*$ the agreement turns out to be far more accurate). 
Note that the anisotropic shear is preserved at very early times and late times in the effective dynamics agreeing respectively with the 
initial and the final value of the anisotropy in the classical solution. However the behavior of directional scale factors changes after 
their bounces, while preserving the anisotropic shear. For this reason, the effective dynamics trajectory (and quantum dynamics) show an 
asymmetric bounce in the sense that the classical solution matched to quantum dynamics before the bounce is different from after the bounce. 
Such a behavior has been confirmed in various anisotropic spacetimes in LQC 
\cite{kevin-chiou,chiou-bianchi3,b1-madrid2,tcpsvt9,bgps-kasner,bgps-inflation,corichi-bianchi1,corichi-bianchi2,
corichi-bianchi3,ks-constant,cs-schw}. Let us also note that the quantum dispersions remain bounded throughout the evolution and take smaller 
values near the bounce. A sharply peaked state chosen at initial times, retains its features throughout the evolution.

Results from another simulation are shown in Fig. \ref{fig:lnv2lnv3_b1_250}. We have plotted the behavior 
of expectation values of relational observables $\widehat{\ln(v_2)|_{b_1}}$ and $\widehat{\ln(v_3)|_{b_1}}$ in relational time $b_1$. Singularity resolution is evident in these plots which show a non-singular bounce 
for $\ln(v_2)$ and a smooth evolution for $\ln(v_3)$. In comparison to the simulation in Fig. \ref{fig:lnv2_lnv1}, the effective dynamics provides a far more more accurate 
approximation to the underlying quantum dynamics. In fact, the effective trajectory sits on the mean value of the quantum curve at all the times for both the relational observables 
shown in this figure. For the same simulation, we have also plotted the behavior of expectation values of $\widehat{\ln(v_3)|_{b_1}}$ versus those of $\widehat{\ln(v_2)|_{b_1}}$ in Fig. \ref{fig:v3v2_250}.  
The non-singular bounce of $\ln(v_3)$ with respect to $\ln(v_2)$ can be seen. Not surprisingly, as in Fig. \ref{fig:lnv2lnv3_b1_250}, the effective trajectory (shown by the solid curve) agrees with the quantum dynamics extremely well at all the scales. The precise agreement between the effective dynamics and the quantum theory is found to be true for all sharply peaked initial states with larger values of $\omega_2^*$ (for the same value of $\omega_3^*$). This seems to imply a validity of effective dynamics at least for large values of $\omega_2^*$ when $\omega_3^*$ is fixed to be large. However, does this conclusion change when smaller values of $\omega_2^*$ are considered? We answer this in the following.

\begin{figure}[b!]
 \includegraphics[angle=0,width=0.5\textwidth]{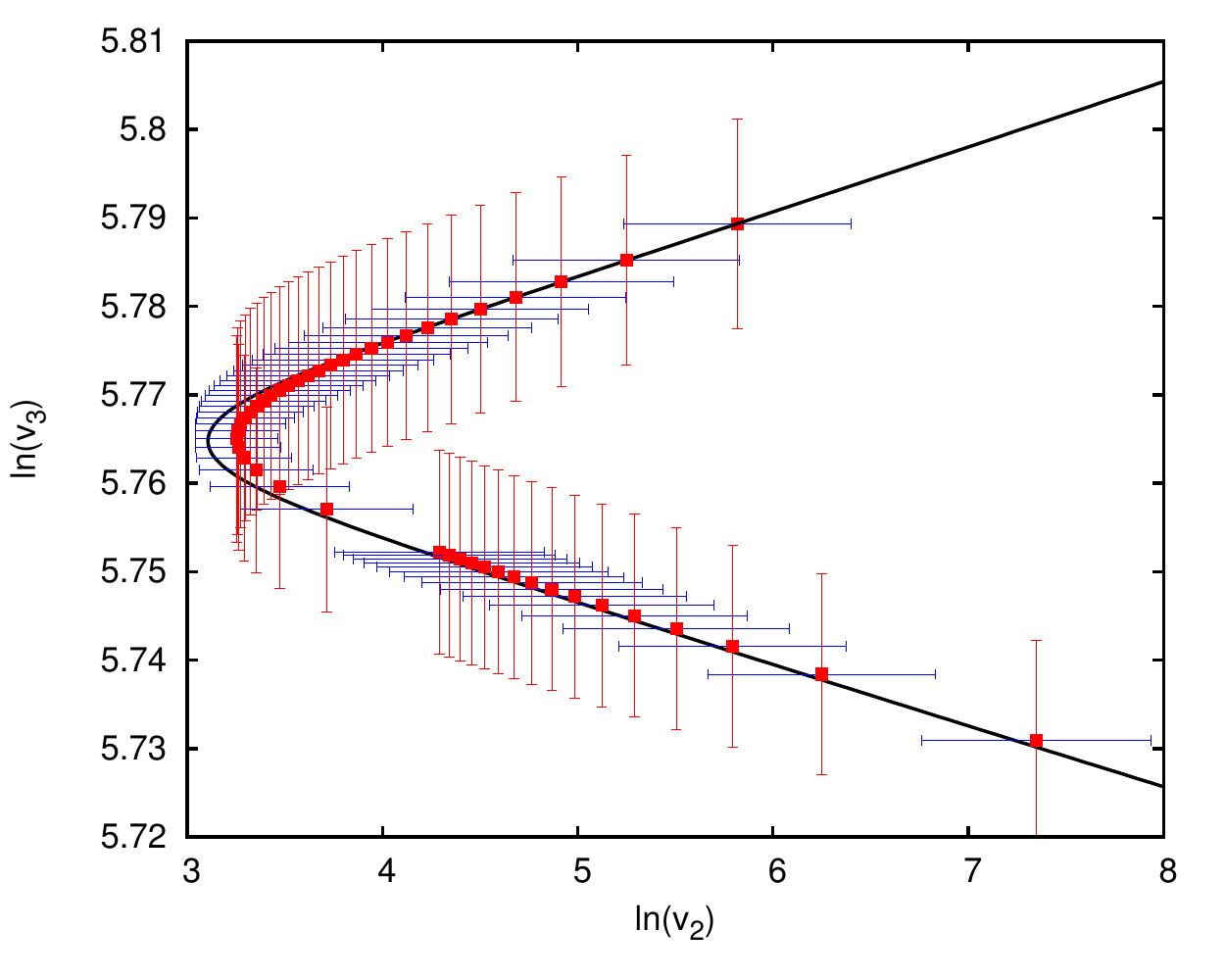}
 \caption{The plot shows the case of  $\omega_2^* = 100$ with $\sigma_2 = 11$, for $\omega_3^* = 1000$ with $\sigma_3 = 40$.    The agreement of the effective dynamics with quantum theory is good, but not as good as in Fig. \ref{fig:v3v2_250}.  For a better visual distinction from the effective theory, quantum expectation values are denoted by thick (red)  points. The dispersion in expectation values of $\widehat{\ln(v_3)|_{b_1}}$ is divided by 10 to fit in the figure.}
\label{fig:v3v2_100} 
\end{figure}

\begin{figure}[t!]
 \includegraphics[angle=0,width=0.5\textwidth]{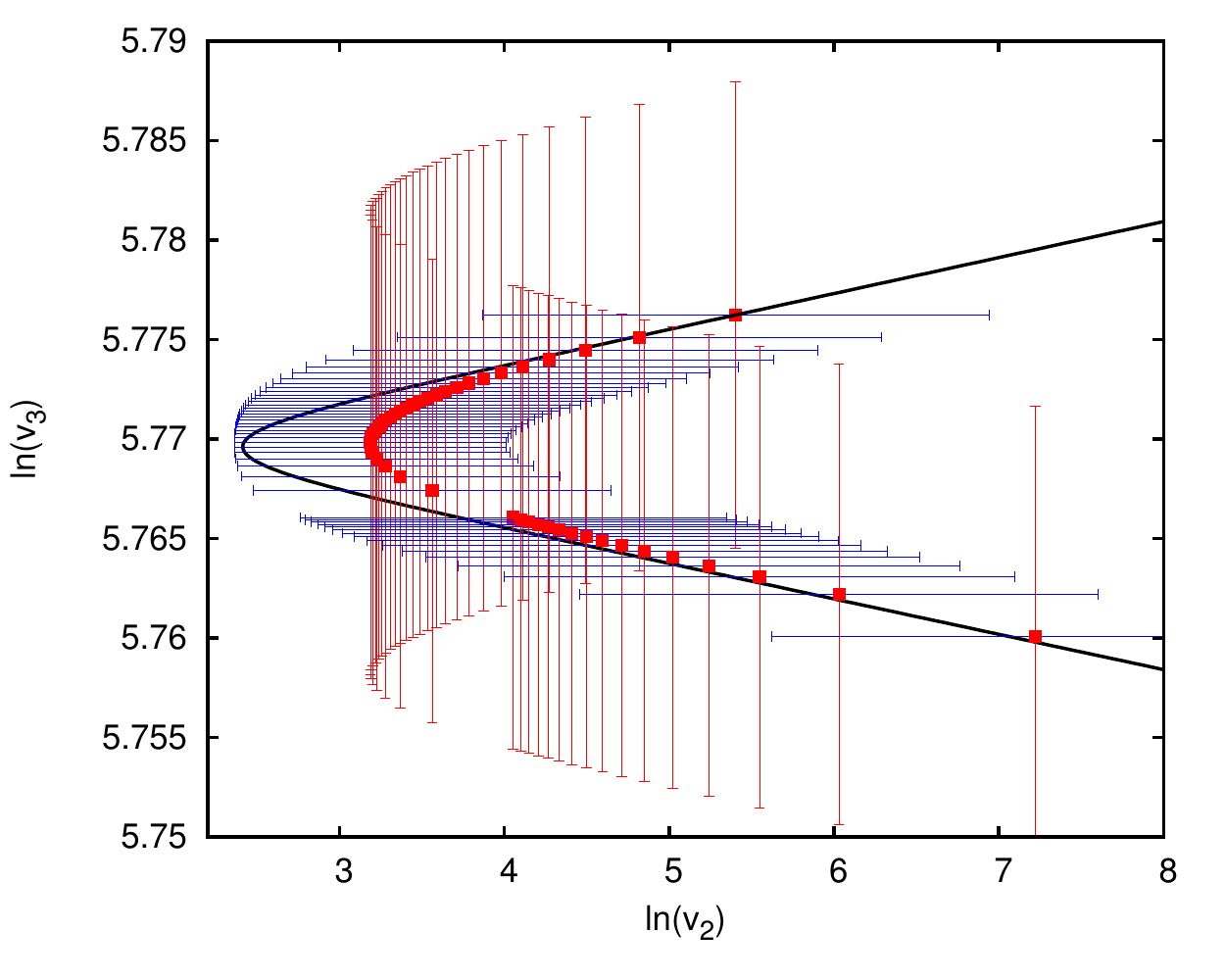}
 \caption{The expectation values for the relational observables for $\ln(v_3)$ and $\ln(v_2)$ are compared with the effective trajectory (solid curve). The parameters are 
 $\omega^*_2 = 50$ with $\sigma_2 = 4$, and other parameters remaining same as in Fig. \ref{fig:v3v2_100}.} \label{fig:v3v2-50}
 \end{figure}

\subsection{Evidence of departure of effective dynamics from quantum theory}

It is evident from Figs. \ref{fig:lnv2_lnv1} and  \ref{fig:lnv2lnv3_b1_250}  that decreasing $\omega_2^*$ for the same value of $\omega_3^*$ causes a slight departure 
between the quantum theory and the effective dynamics. The effective trajectory for $\omega_2^* = 100$ case is such that it predicts bounce at a smaller value of $\ln(v_2)$ and $\ln(v_1)$ in comparison to the one for $\omega_2^* = 250$. The same can be seen to be true if we plot $\ln(v_3)$ versus $\ln(v_2)$. In Fig. \ref{fig:v3v2_100}, a slight departure between the quantum dynamics and effective trajectory (shown by the solid curve) is clearly visible, which is absent for the case of Fig. \ref{fig:v3v2_250}. Comparing these two figures, we find that the effective theory predicts 
smaller bounce volume than the mean value obtained from the quantum dynamics for smaller $\omega_2^*$. It is to be noted that the slight departure from the quantum theory is more significant in the bounce regime. Away from the bounce regime, the agreement of the effective theory with the quantum dynamics is excellent for the simulation in Fig. \ref{fig:v3v2_100} as is the case for the simulation in Fig. \ref{fig:v3v2_250}.

To understand the above trend, we performed various simulations for the smaller values of $\omega_2^*$ keeping rest of the parameters fixed. An example of such a simulation is 
shown in Fig. \ref{fig:v3v2-50} for $\omega_2^* = 50$. In comparison to the simulations in Figs. \ref{fig:v3v2_250} and \ref{fig:v3v2_100}, the quantum state probes deeper Planck regime. This plot of expectation values of $\widehat{\ln(v_3)|_{b_1}}$ and $\widehat{\ln(v_2)|_{b_1}}$ clearly shows that there is an increased 
departure of the effective dynamics from the quantum dynamics in this case. Though the effective curve is still within the dispersions of the relational observables, it underestimates  the bounce volume in the quantum theory significantly. This implies that bounce in the quantum theory occurs at smaller spacetime curvature than is estimated from the effective dynamics. Two things are notable in this simulation. First that even though for the effective dynamics there are departures from the quantum theory in the bounce regime, away from the bounce regime effective dynamics is an excellent agreement with the quantum theory. 
Further, for this particular state, fluctuations are quite high. (As in the Fig. \ref{fig:v3v2_100}, in this figure the dispersion in  $\widehat{\ln(v_3)|_{b_1}}$ is divided by 10 for a better visualization). This simulation confirms that the bounce occurs for states which are not necessarily sharply peaked. As in the previous cases, fluctuations remain bounded throughout the evolution and become smaller in the bounce regime. 

\begin{figure}[b!]
 \includegraphics[angle=0,width=0.5\textwidth]{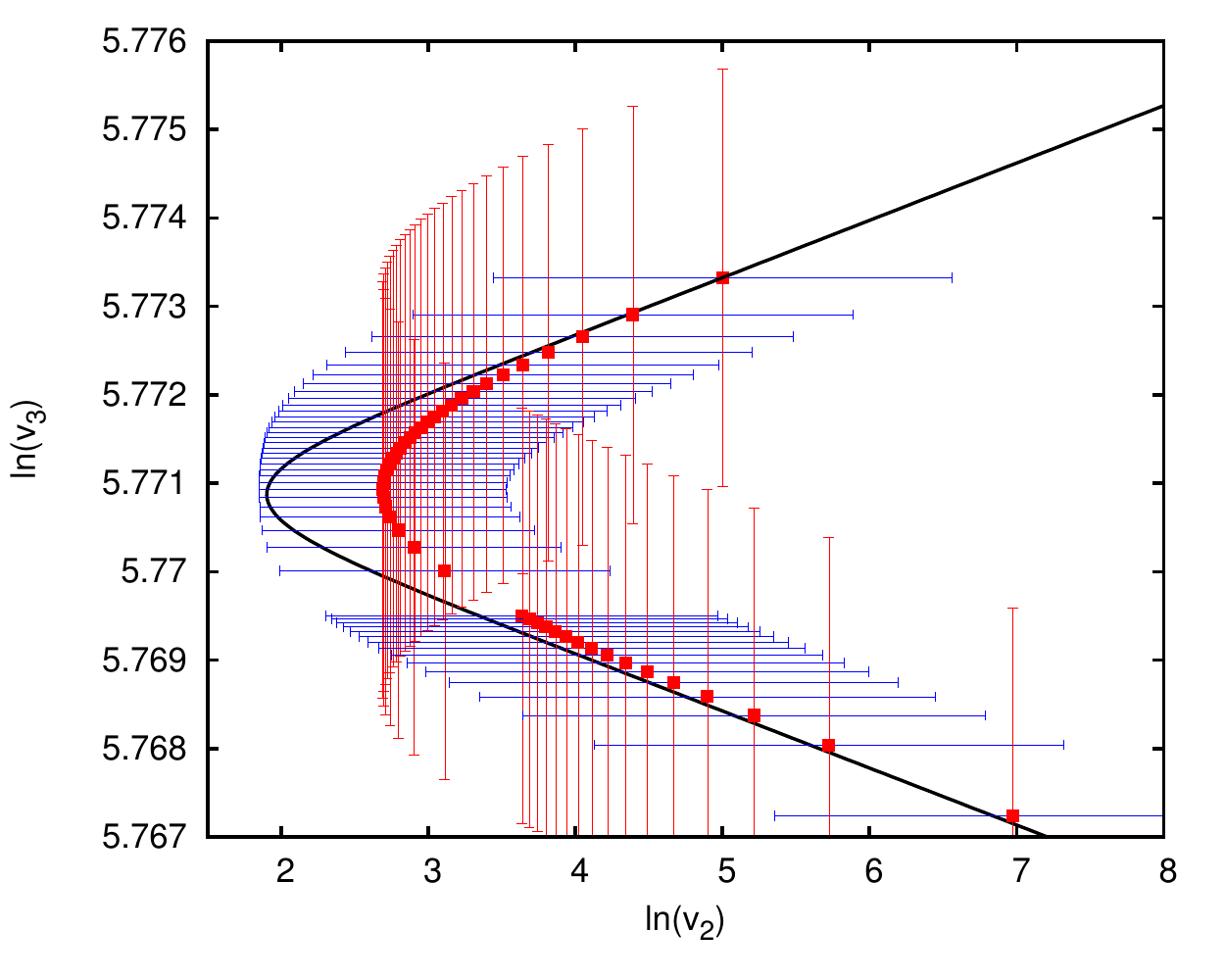}
 \caption{This plot shows the results from the simulation for  $\omega^*_2 = 30$ with $\sigma_2 = 4$ (right plot). In contrast to simulations in Fig. \ref{fig:v3v2_100}, effective dynamics (solid black curve) is not a good approximation to the quantum dynamics (thick red dots), especially in the bounce regime.   }
 \label{fig:v3v2_30}
\end{figure}

\begin{figure}[t!]
 \includegraphics[angle=0,width=0.48\textwidth]{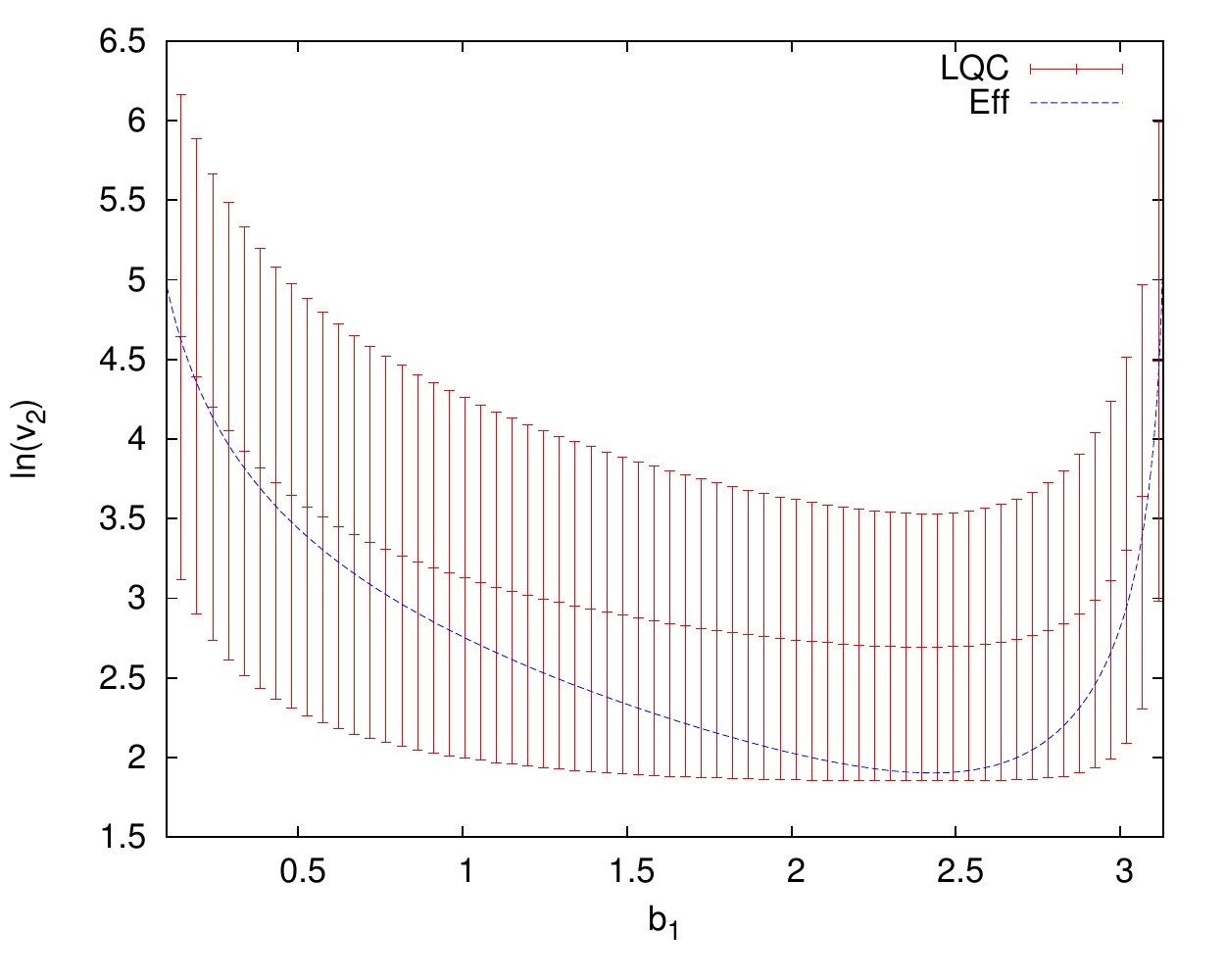}
 \includegraphics[angle=0,width=0.48\textwidth]{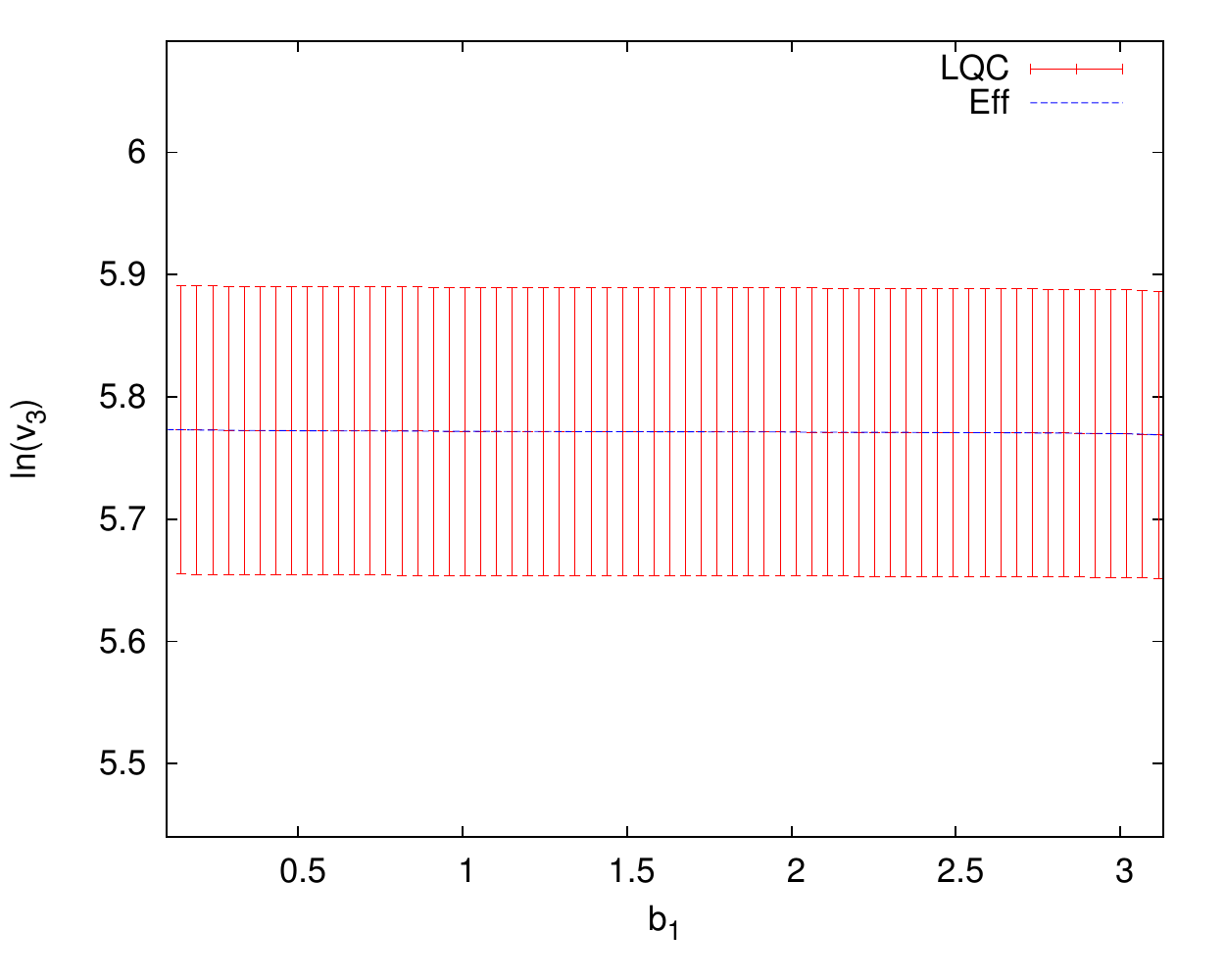}
\caption{The behavior of expectation values of relational observables $\widehat{\ln(v_2)|_{b_1}}$ and $\widehat{\ln(v_3)|_{b_1}}$ is shown for 
$\omega_2^* = 30$ with $\sigma_2 = 4$. Here $\omega_2^* = 1000$ with $\sigma_2 = 40$. In contrast to the simulation in Fig. \ref{fig:lnv2lnv3_b1_250}, we see significant departures between 
the effective and the quantum trajectories. Other parameter values are same as in Fig. \ref{fig:lnv2lnv3_b1_250}. Effective dynamics 
predicts smaller bounce volumes than the mean values of the observables in the quantum theory.}
\label{fig:lnv2lnv3_b1_30}
 \end{figure}

Let us now consider the case of one of the extreme cases studied in our analysis, shown in Fig. \ref{fig:v3v2_30}.  In this one of most computationally expensive simulation, we have considered $\omega_2^* = 30$. This state probes the deepest possible quantum regime in our analysis, and confirms the expectations from Fig. \ref{fig:v3v2-50}. The effective dynamics has a significant departure from the mean values in the quantum theory in the bounce regime. Away from the bounce regime, the effective dynamics again provides a good approximation to the quantum dynamics. The fluctuations for this simulation are quite high, and the dispersions in the expectation value of $\widehat{\ln(v_3)|_{b_1}}$ are divided by 50 to fit in the figure. 
Nevertheless, effective trajectory always lies within the dispersions of the observables. As in the case of the other simulations showing departures from quantum dynamics, effective theory underestimates the volume at the bounce. For the same simulation, we have plotted in Fig. \ref{fig:lnv2lnv3_b1_30} the behavior of expectation values of $\widehat{\ln(v_2)|_{b_1}}$ and $\widehat{\ln(v_3)|_{b_1}}$ in time $b_1$. Various features become clear from this plot which shows a non-singular evolution of the above relational observables in time. In contrast to the simulation for $\omega_2^* = 250$ in Fig. \ref{fig:lnv2lnv3_b1_250}, there is a significant departure of effective dynamical trajectory from the quantum theory for the relational observable $\ln(v_2)$. In case of the $\ln(v_3)$ the results do not change and there is little departure between the quantum theory and effective dynamics. This is not surprising because 
the difference between the two simulations is only in the values of $\omega_2^*$. As we can see, the effective trajectory is almost at the values of maximum dispersion for $\widehat{\ln(v_2)|_{b_1}}$ in the bounce regime. Further, the bounce time in the case of effective and quantum dynamics is visibly different in this simulation, which is not the case for the 
simulation in  Fig. \ref{fig:lnv2lnv3_b1_250}.

These simulations indicate that states with different $\omega_2^*$ which have same value of $\omega_3^*$, the departure between quantum theory and effective dynamics 
increases as $\omega_2^*$ is decreased. Of course with limited simulations discussed so far, it is difficult to find any subtle features in this relationship. However, somethings become concretely clear. States which probe deep Planck regime bounce at volumes greater than those predicted by the effective theory. Such states also have larger fluctuations. And in a way these fluctuations help in quantum repulsiveness causing bounces to occur at smaller spacetime curvature than one would expect from the effective dynamics. 
The situation is in harmony  to the results earlier obtained for the isotropic model in LQC \cite{numlsu-2,numlsu-3}. There it was shown that states which bounce closer to the classical big bang singularity provide largest departures between quantum theory and effective dynamics. As in the present analysis, the effective dynamics overestimates the spacetime curvature at the quantum bounce.

Thus, we have so far established that irrespective of  the value of $\omega_2^*$, singularity resolution occurs. For states with very small $\omega_2^*$ there are departures between quantum theory and effective dynamics. For reasonable values of $\omega_2^*$, effective dynamics provides an excellent approximation to the quantum theory. In the following we quantify the departure between the quantum theory and effective dynamics for various simulations performed in our analysis.

 \subsection{Departure of effective dynamics from quantum theory: quantitative aspects}
For the simulations discussed so far we have found that as $\omega_2^*$ is decreased keeping $\omega_3^*$ fixed then the departure between the effective dynamics 
and quantum theory seems to increase. This departure appears to be maximum near the bounce. In order to understand the way this departure depends on various 
parameters, we focus on just one relational observable which is expectation values of $\widehat{\ln(v_2)|_{b_1}}$. To extract the departure we find the difference between the 
the expectation value of this observable at the bounce in the quantum theory and its analog value in the effective dynamics. This difference, for sharply peaked states, is a measure of the relative difference in the bounce volume in the quantum theory and effective dynamics. We understand the dependence of this departure on the following parameters: (i) the value of $\omega_2^*$ for different values of  absolute fluctuation $\sigma_2$, (ii) the value of $\sigma_2$ for various values of $\omega_2^*$, (iii) the relative fluctuation in $\omega_2^*$, given by $\sigma_2/\omega_2^*$, and (iv) the dispersion in the relational observable $\widehat{\ln(v_2)|_{b_1}}$.

\begin{figure}[!t]
 \includegraphics[angle=0,width=0.5\textwidth]{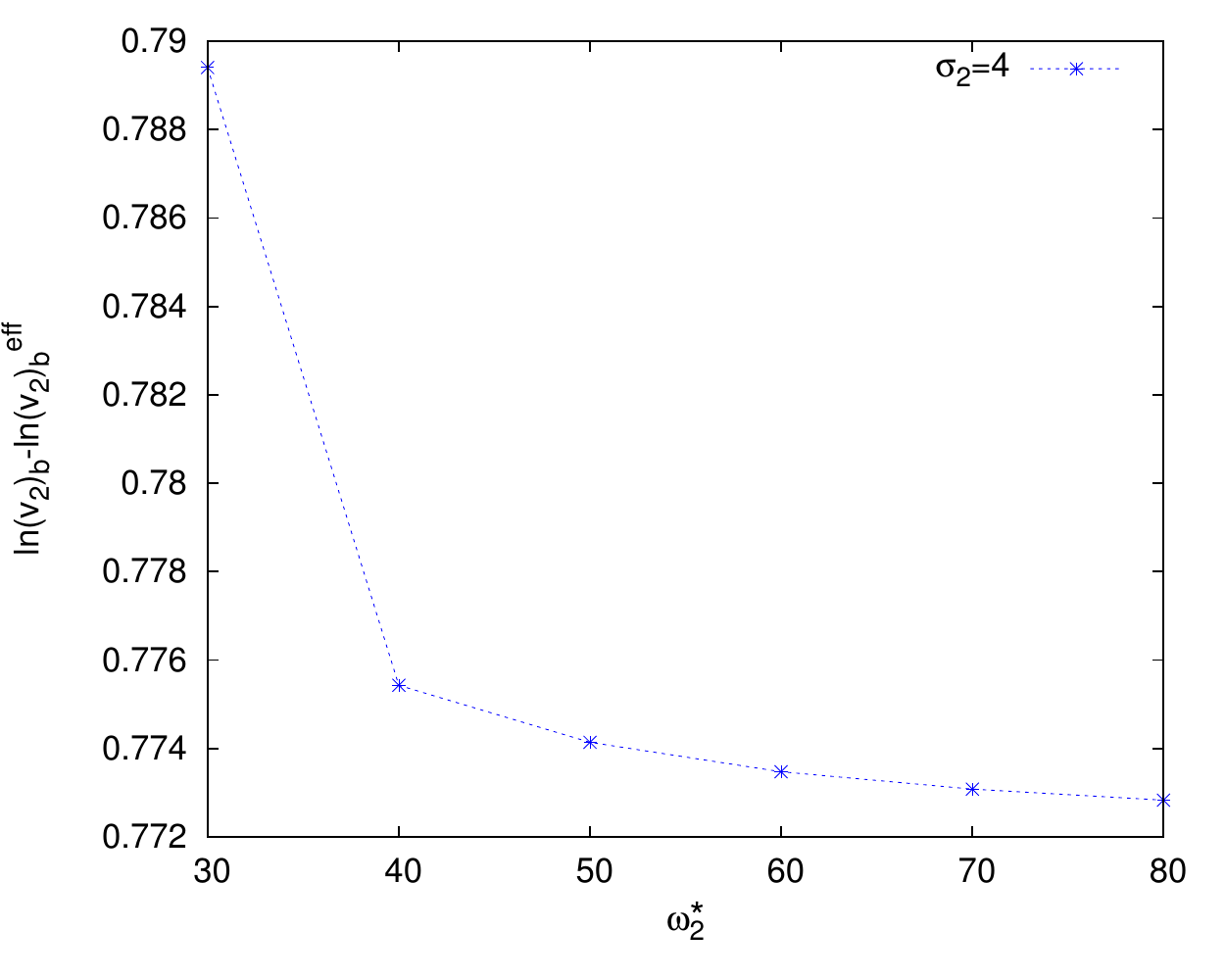}
 \caption{The plot shows the difference in the value of relational observable corresponding to $\ln(v_2)$ at the bounce in the quantum theory and the same in the effective dynamics as a 
function of $\omega_2$ for $\sigma_2 = 4$.}
 \label{fig:deltaeff_vs_w2_s2_4}
\end{figure}

\begin{figure}[b!]
 \includegraphics[angle=0,width=0.5\textwidth]{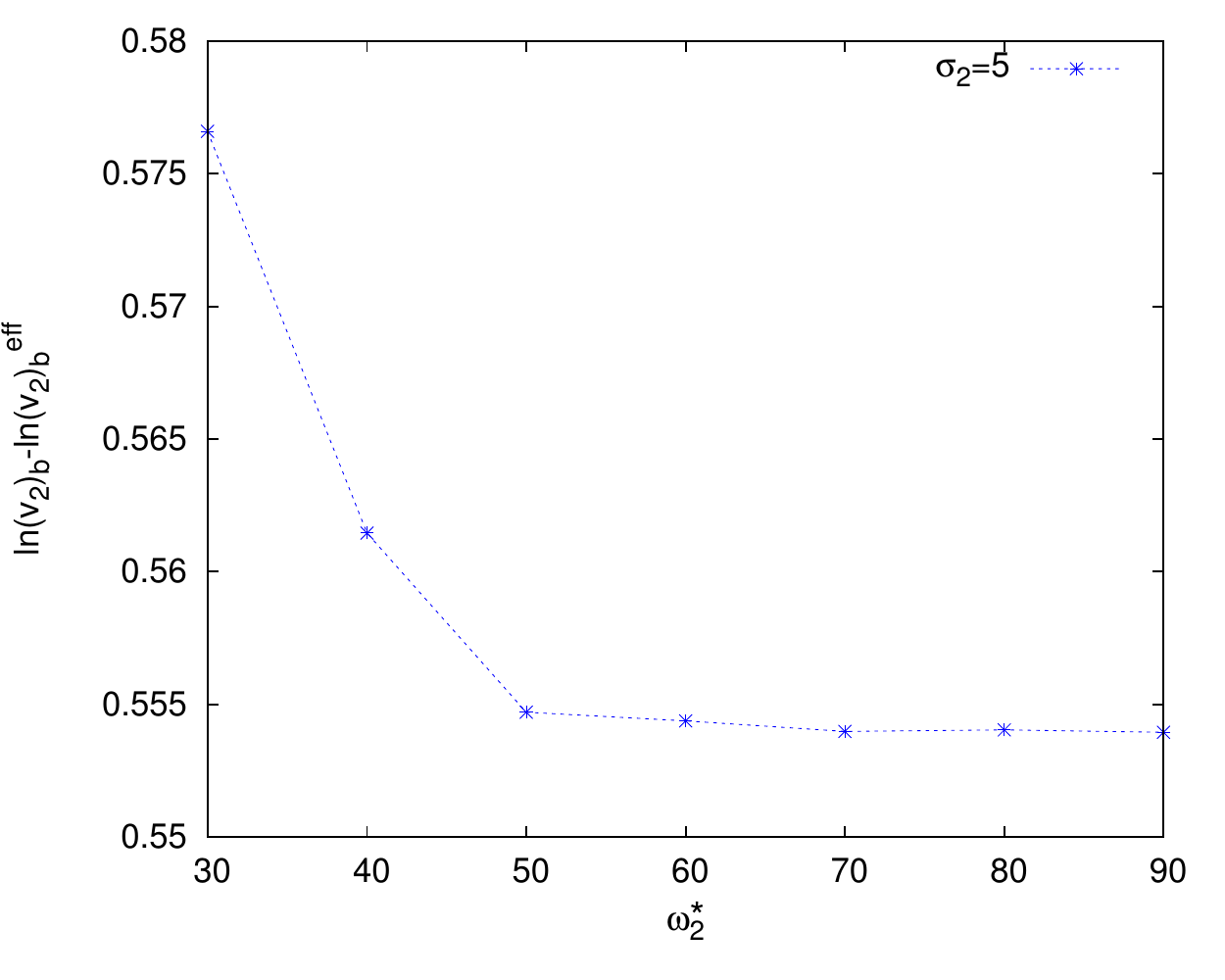}
 \caption{Difference between the expectation values of $\widehat{\ln(v_2)|_{b_1}}$, and $\ln(v_2)$ in effective 
theory at the bounce is shown for different values of $\omega_2$ for $\sigma_2 = 5$.}
 \label{fig:deltaeff_vs_w2_s2_5}
\end{figure}

Continuing with our findings in Sec. IVB, let us analyze the departure between the quantum theory and the effective dynamics by studying the way it changes when $\omega_2^*$ is changed. We keep the absolute fluctuation $\sigma_2$ and all other parameters including $\omega_3^*$ fixed. Note that for any given value of $\sigma_2$, it is possible to explore only a limited range of $\omega_2^*$. In Fig. \ref{fig:deltaeff_vs_w2_s2_4}, we have shown the variation of this difference for the case of $\sigma_2 = 4$. In this range of simulations from $\omega_2^*=30$ till $\omega_2^*=80$ for $\omega_3^*=1000$, we find that the departure between the quantum theory and effective theory slowly increases as $\omega_2^*$ is decreased till $\omega_2^* = 50$, but the departure quickly increases at around $\omega_2^* = 40$ and becomes much larger at $\omega_2^*=30$.  Another set of simulations, this time for $\sigma_2 = 5$ are shown in Fig. \ref {fig:deltaeff_vs_w2_s2_5}. The range of $\omega_2^*$ in these simulations is 
similar. 
We find that for the values greater than or equal to $\omega_2^* = 50$, the departure between the effective theory and quantum dynamics approximately increases slowly. For smaller values of $\omega_2^*$ 
there is a rapid increase with the largest value of departure at the smaller $\omega_2^*$ probed in this set of simulations. 
Figs. \ref{fig:deltaeff_vs_w2_s2_4} and \ref{fig:deltaeff_vs_w2_s2_5} thus show that the largest departure between the effective dynamics and quantum theory appears at smallest value of $\omega_2^*$. For other values of $\omega_2^*$, both sets of simulations show an almost monotonic increase in the value of departure as $\omega_2^*$ is decreased.  

\begin{figure}[!t]
 \includegraphics[angle=0,width=0.5\textwidth]{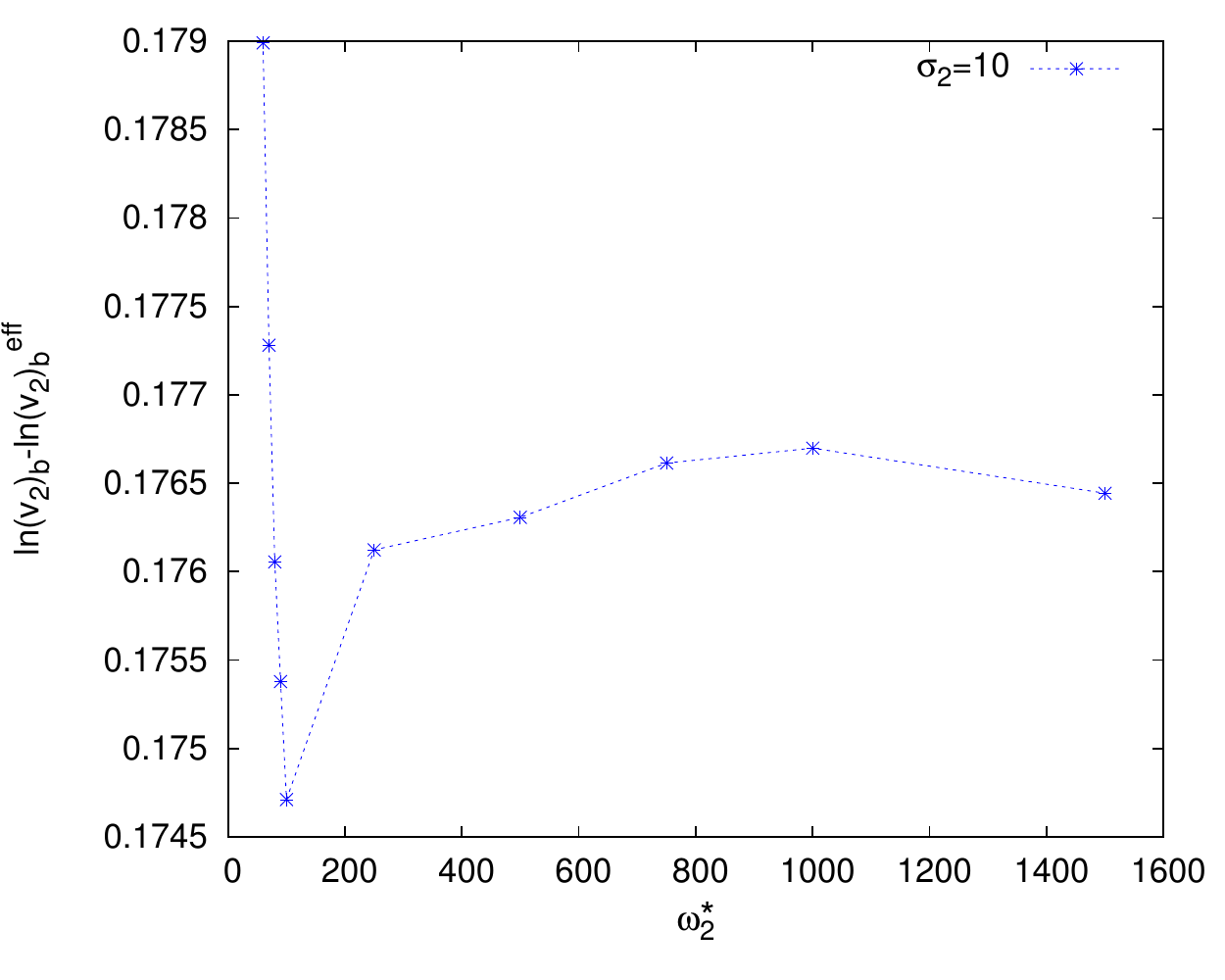}
 \caption{The expectation values of $\widehat{\ln(v_2)|_{b_1}}$ at bounce are compared with their counterpart in the effective theory. The 
value of $\sigma_2 = 10$ is chosen for different values of $\omega_2$.}
 \label{fig:deltaeff_vs_w2_s2_10}
\end{figure}

\begin{figure}[b!]
 \includegraphics[angle=0,width=0.5\textwidth]{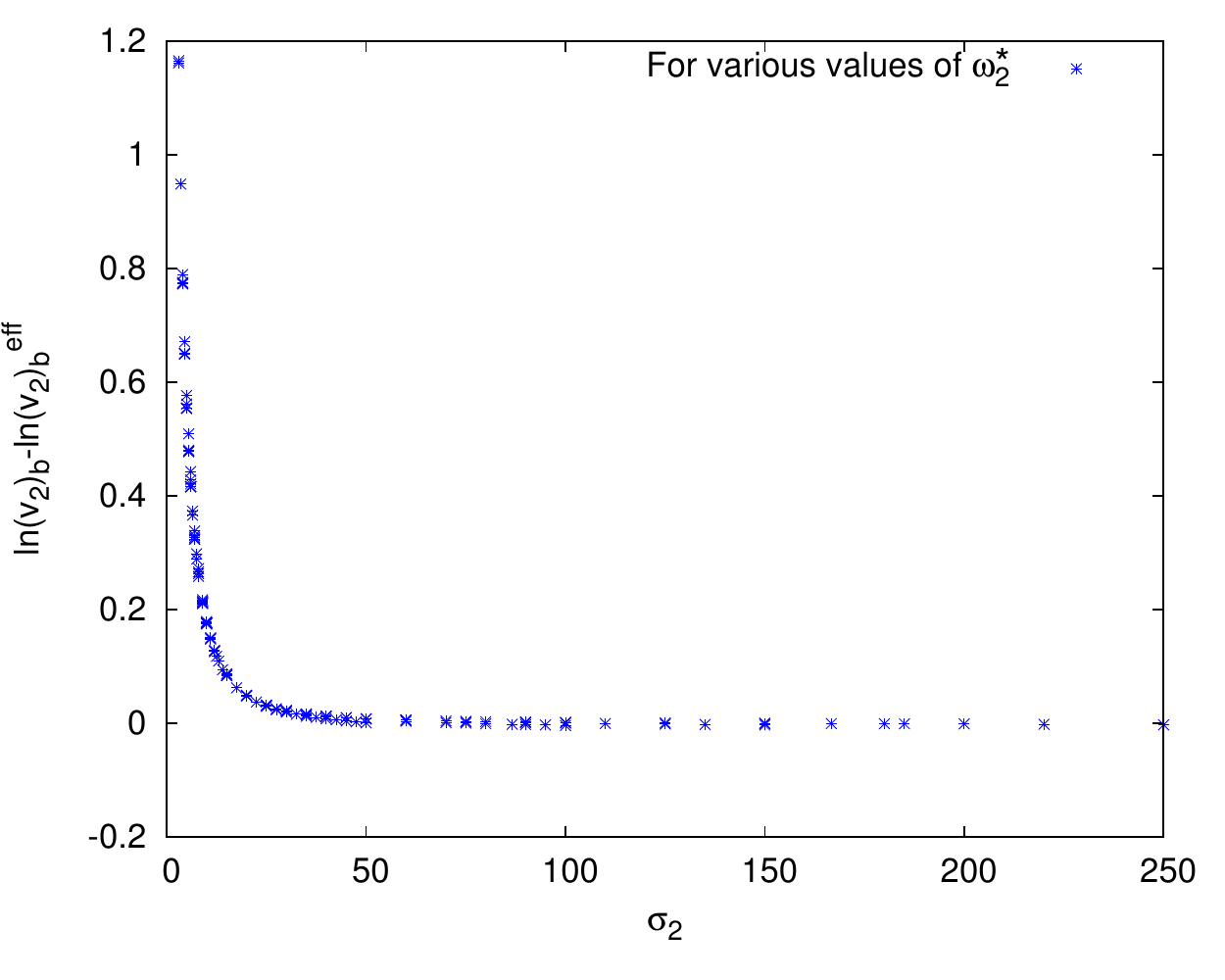}
 \caption{Variation of the difference between the expectation values of $\widehat{\ln(v_2)|_{b_1}}$ and corresponding values in the effective theory at the bounce is plotted versus $\sigma_2$ for various values of $\omega_2$.}
 \label{fig:all_deltaeff_vs_s2}
\end{figure}

The above monotonic behavior is not found to hold  for simulations with a larger range of $\omega_2^*$ which allow a larger value of $\sigma_2$. One such set of simulations is 
shown in Fig. \ref{fig:deltaeff_vs_w2_s2_10}, where $\omega_2^*$ ranges from 60 till 1500. Some distinguishing features are evident in comparison to Figs. \ref{fig:deltaeff_vs_w2_s2_4} 
and \ref{fig:deltaeff_vs_w2_s2_5}. Unlike the simulations in the latter figures, for the larger values of $\omega_2^*$ there is no slight increase in 
departure between quantum and effective dynamics as $\omega_2^*$ is decreased. Rather, we find the departure to increase and then decrease. Interestingly, the departure rapidly decreases 
below $\omega_2^* = 250$, becomes minimum at $\omega_2^* = 100$, and then very rapidly increases monotonically. Such an overall non-monotonic behavior which is in striking contrast to 
simulations for $\sigma_2 = 4$ and $\sigma_2 = 5$  was also seen for other simulations in a similar range of $\omega_2^*$ for different values of $\sigma_2$. Nevertheless, in agreement 
with above sets of simulations we always found that irrespective of the choice of $\sigma_2$, the departure between the quantum and effective dynamics increases when $\omega_2^*$ is 
decreased for smaller values of $\omega_2^*$. These simulations show that it is not guaranteed that an increase in the value of $\omega_2^*$, increases the agreement between the quantum theory and effective dynamics.  \\

Though the departure between the quantum theory and effective dynamics does not show a monotonic variation for all values of $\omega_2^*$, interestingly a monotonic variation is 
found for the dependence on the fluctuation $\sigma_2$. In Fig. \ref{fig:all_deltaeff_vs_s2}, we plot the departure between the quantum and effective dynamics at 
the bounce versus $\sigma_2$ for all values of $\omega_2^*$. For larger values of $\sigma_2$, there is a little variation in the difference between the quantum and 
effective dynamics as $\sigma_2$ is varied. Between $\sigma_2 = 50$ till $\sigma_2 = 250$ there is only a little increase as $\omega_2^*$ is decreased. For large values of $\sigma_2$, the difference between the 
quantum and effective dynamics vanishes. For very large value of $\sigma_2$ we find some evidence that the bounce volume for $\widehat{\ln(v_2)|_{b_1}}$ in the quantum theory 
is smaller than the one in the effective theory. On the other hand, we find that for $\sigma_2$ smaller than 15 there is a significantly large increase in departure 
between the values of $\ln(v_2)$ at the bounce in quantum and effective dynamics. As the absolute fluctuation in $\omega_2^*$ decreases to small values, the departure 
becomes very large. Thus, the agreement between  the effective dynamics with the quantum theory is not sensitive to the value of $\sigma_2$ for large values of $\sigma_2$, it is extremely sensitive for small values of $\sigma_2$. \\

\begin{figure}[b!]
 \includegraphics[angle=0,width=0.5\textwidth]{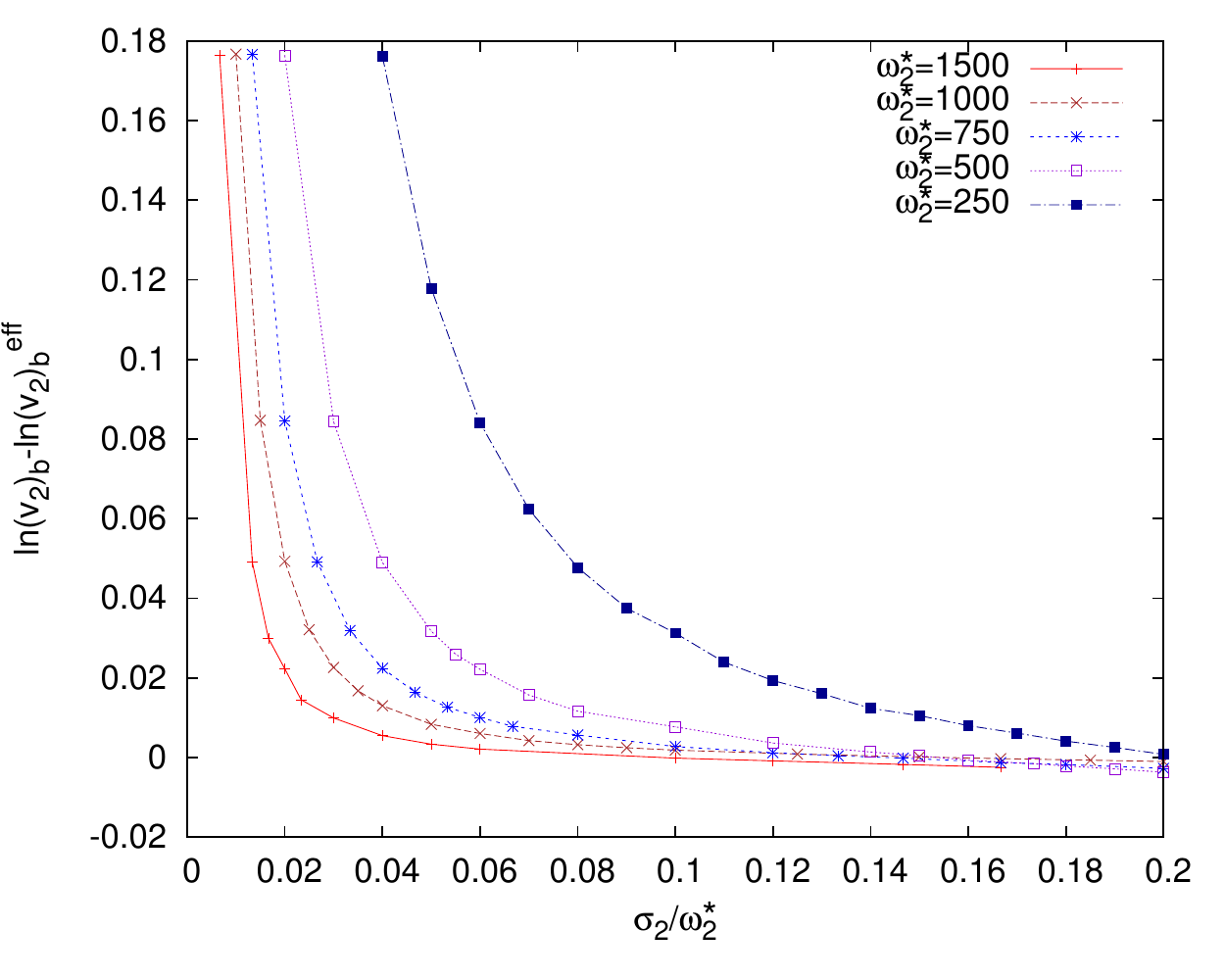}
 \caption{Variation of the departure of the effective dynamics from quantum theory versus relative dispersion in $\omega_2$ is shown for  $\ln(v_2)$ at the bounce. Various simulations for $\omega_2^* \geq 250$ are shown.} \label{fig:large_deltaeff_vs_s2byw2_v2} 
 \end{figure}

The variation of the difference between the predicted bounce volume in the effective dynamics and the expectation value in the quantum theory is again found to be monotonic 
when plotted with respect to the relative fluctuation in $\omega_2^*$. In Fig. \ref{fig:large_deltaeff_vs_s2byw2_v2} we show results from various simulations for all the $\omega_2^* \geq 250$. We find that as relative dispersion $\sigma_2/\omega_2^*$ increases, the effective dynamics becomes a less accurate approximation of the quantum theory. For larger values of $\omega_2^*$, for small relative fluctuations there is a significant increase in the departure between the bounce volumes in quantum theory and effective dynamics. For smaller values of $\omega_2^*$ the change in variation of the difference between quantum and effective dynamics results is not so sharp. As the value of $\sigma_2/\omega_2^*$ becomes very large, 
the departure between the quantum and effective theory vanishes. In the same regime, curves corresponding to different values of $\omega_2^*$ intersect each other. As in the case of Fig. \ref{fig:all_deltaeff_vs_s2}, we find that difference between the logarithm of the directional volume $v_2$ at the bounce in quantum theory with the one in the effective theory becomes negative for 
some simulations in the regime of very large dispersions. Hence for almost all the simulations, effective theory underestimates the directional volume at the bounce but for a few simulations the situation is reversed.

 \begin{figure}[t!]
  \includegraphics[angle=0,width=0.5\textwidth]{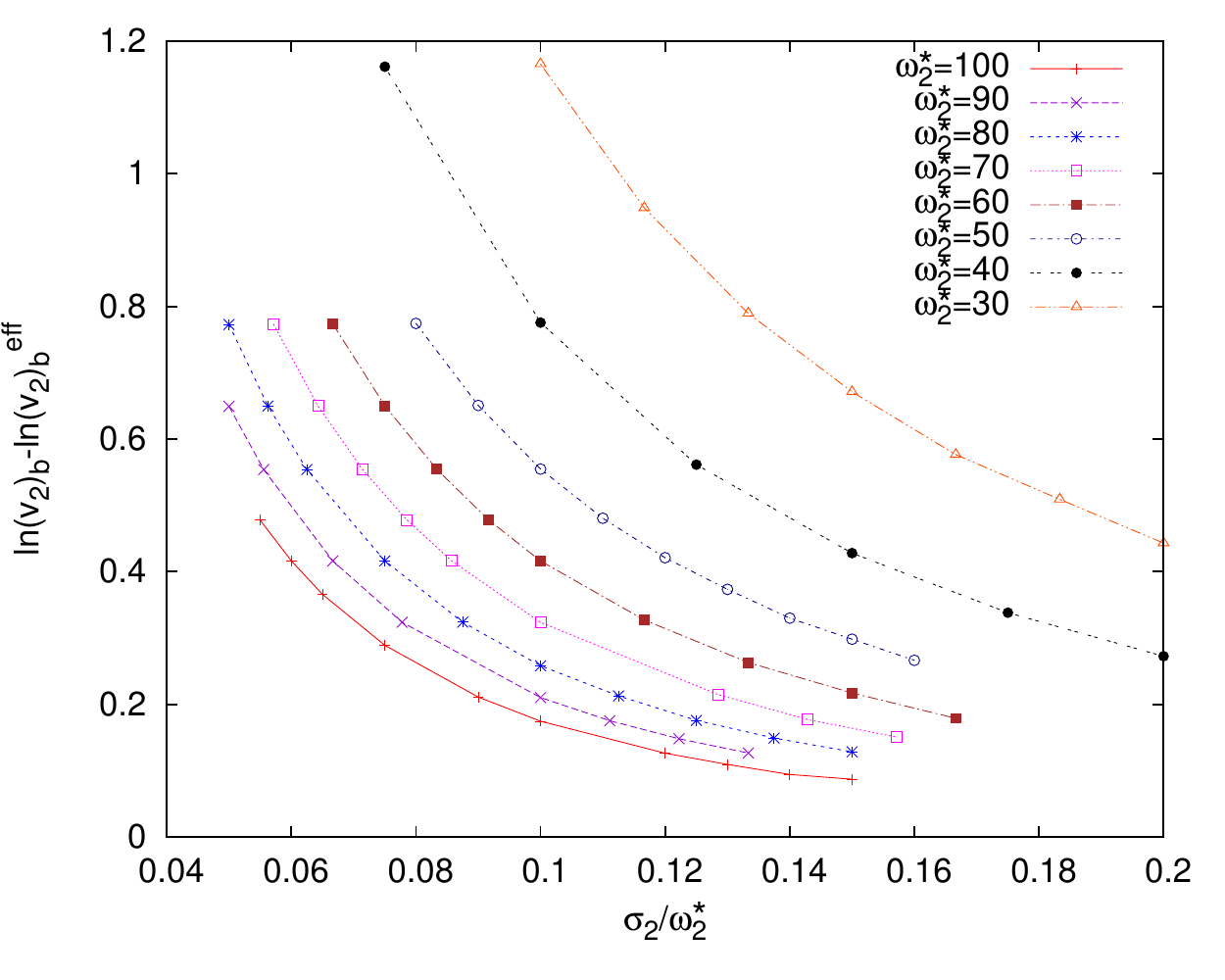}
  \caption{Difference between the values of $\ln(v_2)$ in effective theory at the bounce and the value determined from the expectation values of $\widehat{\ln(v_2)_{b_1}}$ in LQC are shown with respect to $\sigma_2/\omega_2^*$.  
  Various simulations for lower values of $\omega_2^*$ ($\omega_2^* \leq 100$) are shown. }
\label{fig:small_deltaeff_vs_s2byw2_v2}  
\end{figure}

\begin{figure}[b!]
 \includegraphics[angle=0,width=0.5\textwidth]{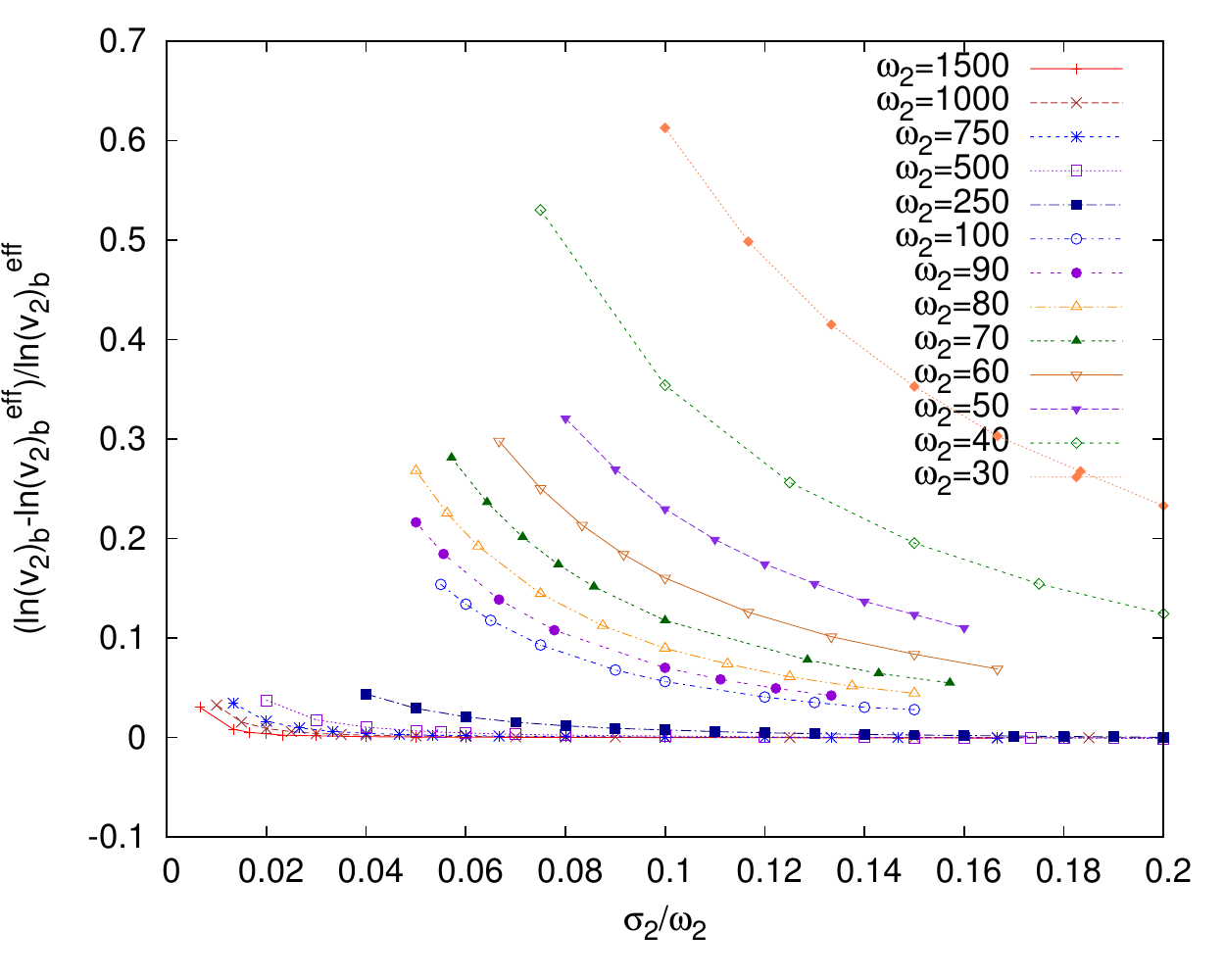}
 \caption{Variation of the relative difference in $\ln(v_2)$ at the bounce for the quantum theory and effective dynamics is plotted versus the relative dispersion in $\omega_2$. } 
 \label{fig:deltaeffrel_vs_s2byw2_v2} 
 \end{figure}

\begin{figure}[!t]
 \includegraphics[angle=0,width=0.5\textwidth]{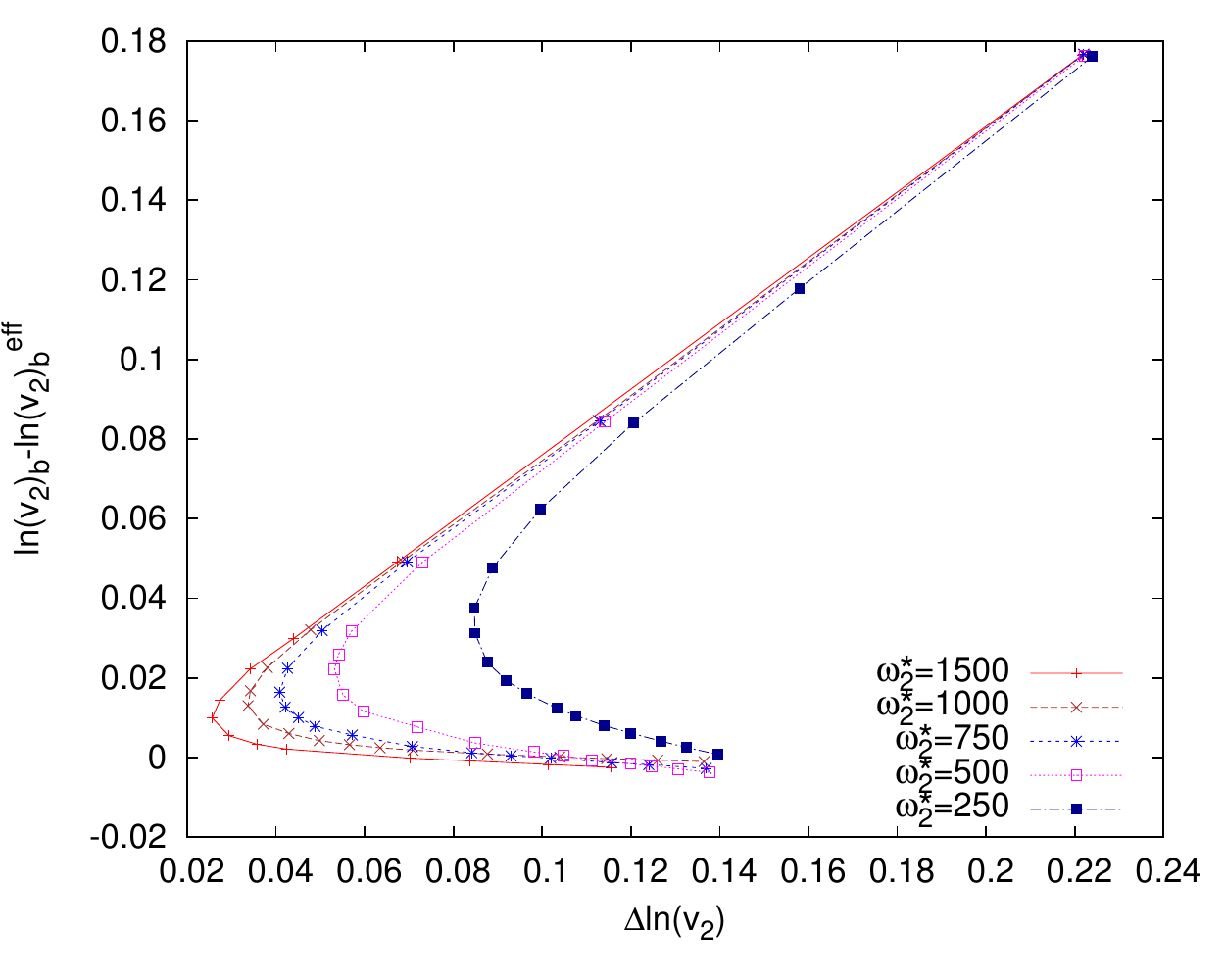}
 \caption{The plot shows the way the difference of the bounce volume in quantum and effective dynamics changes with respect to $\omega_2^*$. A turnaround and a non-monotonic behavior is present for each value of $\omega_2^*$.}\label{fig:largeturnaround}
 \end{figure}

\begin{figure}[!b]
 \includegraphics[angle=0,width=0.5\textwidth]{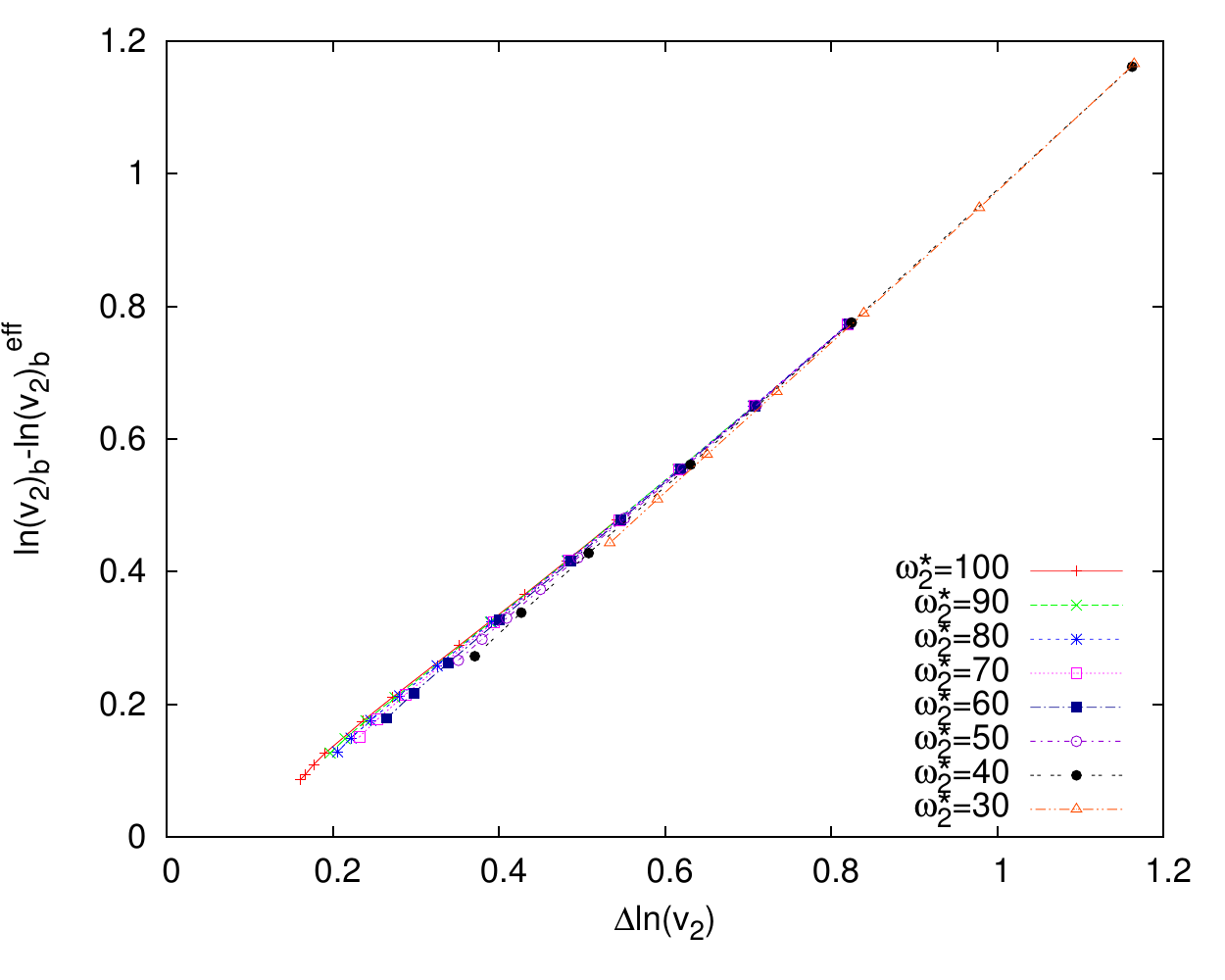}
\caption{Departure between the expectation values of $\widehat{\ln(v_2)|_{b_1}}$ and corresponding values in the effective theory at the 
bounce is plotted versus absolute dispersion of the former. Various simulations for $\omega_2^*$ between 30 and 100 are shown. }
\label{fig:smallturnaround}
 \end{figure}

In Fig. \ref{fig:small_deltaeff_vs_s2byw2_v2}, various simulations for the smaller values of $\omega_2^*$ are plotted versus $\sigma_2/\omega_2^*$. The behavior of the difference between 
the value at the bounce of the relational observable for $\ln(v_2)$ in the quantum theory and its analog in the effective dynamics follows the trend we found for states with larger $\omega_2^*$.
At larger values of  $\sigma_2/\omega_2^*$, the departure is smaller irrespective of the value of $\omega_2^*$. As the value of relative fluctuation decreases, the difference between the quantum and effective dynamics increases. This increase is rapid for smaller values of $\sigma_2/\omega_2^*$. Due to the large spreads  associated with these states, it is difficult to explore a much wider range of parameter space. For this reason the intersection of the curves as seen for the case of larger 
$\omega_2^*$ is not yet visible in the plot. Note that in comparison to the simulations for larger $\omega_2^*$, the difference between quantum and effective dynamics is already more 
significant even for this range of parameters. We expect this difference to further increase substantially for further smaller values of relative fluctuations in $\omega_2^*$.

Another useful parameter to understand the departure between the quantum theory and the effective dynamics is the relative difference in the value of observable $\ln(v_2)$. In Fig. \ref{fig:deltaeffrel_vs_s2byw2_v2}, we plot this 
difference with respect to the relative fluctuation $\sigma_2/\omega_2^*$ for all the values of the $\omega_2^*$. For larger values of $\omega_2^*$ we find that the relative difference in bounce volume is smaller and approaches zero quickly 
as the relative fluctuation $\sigma_2/\omega_2^*$ increases. On the other hand, for simulations with $\omega_2^* \lesssim 100$, the relative difference is significantly larger. It decreases as the relative fluctuation in $\omega_2^*$ increases but 
can remain substantial even for the largest studied relative fluctuation $\sigma_2/\omega_2^*$ in our simulations. For the simulation corresponding to $\omega_2^* = 30$, the relative difference is approximately 50\% at the largest studied 
val;ue of relative fluctuation. This plot suggests that for different values of $\omega_2^*$ there is some sort of attractor behavior. For larger values of 
relative fluctuation in $\omega_2^*$, the relative difference in logairthm of bounce volume vanishes. One can similarly study  behavior of the latter parameter when 
versus the fluctuations $\sigma_2$. However, such a plot results in a very similar curve as in Fig. \ref{fig:all_deltaeff_vs_s2} whose implications are already discussed above. \\

So far we have investigated the relationship between the departure of effective dynamics from quantum theory for the bounce volume in $\ln(v_2)$ in terms of $\omega_2^*$ and its fluctuations. We now study the way this departure depends on the fluctuations in the expectation values of  $\widehat{\ln(v_2)|_{b_1}}$. This is a measure of the relative fluctuation in the corresponding volume observable. In Fig. \ref{fig:largeturnaround}, we have plotted results from various simulations for $\omega_2^* \geq 250$. We find an interesting behavior that  the departure of the effective theory from quantum dynamics decreases as the dispersion in $\widehat{\ln(v_2)_{b_1}}$ decreases but this trend stops below a certain value of dispersion which is determined by the value of $\omega_2^*$. This corresponds to a turnaround in the behavior of the departure between the quantum and effective dynamics.  Below this turnaround, the departure from quantum dynamics only decreases on increasing the dispersion $\Delta \ln(v_2)_{b_1}$ 
of the state. Note that the part of the curves at the top right of the turnaround correspond to smaller values of fluctuations in $\omega_2^*$, where as the bottom  right part corresponds to larger fluctuations in $\omega_2^*$. Noting this, the above behavior can then be restated as follows. As the dispersion in the relational observable corresponding to $\ln(v_2)$ decreases to a certain value, the departure of the effective dynamics from quantum dynamics can increase or decrease depending on whether the state corresponds to data points approaching the turnaround in curves in Fig. \ref{fig:largeturnaround} from bottom (larger dispersions in $\omega_2^*$) or from top (smaller dispersions in $\omega_2^*$). We find that for any given value of $\omega_2^*$ there is a minimum allowed value of $\Delta \ln(v_2)_{b_1}$. This non-monotonic behavior is found for all value of $\omega_2^*$. The turnaround point in the behavior of 
the difference between quantum and effective dynamics occurs follows an inverse relationship between the value of $\omega_2^*$ and $\Delta \ln(v_2)_{b_1}$. For larger values of $\omega_2^*$, the turnaround of the behavior of difference occurs at smaller values of dispersion. Finally, let us note that as in the case Figs. \ref{fig:all_deltaeff_vs_s2} and \ref{fig:large_deltaeff_vs_s2byw2_v2}, we find that for some simulations the difference between bounce volume in quantum and effective theory becomes negative. This occurs in the regime where different curves converge and intersect.

In Fig. \ref{fig:smallturnaround} we study the dependence of the departure between quantum and effective dynamics on dispersion in expectation values of $\widehat{\ln(v_2)|_{b_1}}$ for smaller values of $\omega_2^*$. The behavior 
of the curves captures the characteristics of the curves at the top right part in Fig. \ref{fig:largeturnaround}. It is to be noted that the simulations on the bottom right part of 
the curves in Fig. \ref{fig:largeturnaround} are  computationally most demanding, especially for smaller values of $\omega_2^*$. Only for this reason, the non-monotonic behavior in the 
difference of quantum and effective dynamics in relation to dispersion in volume is not visible in Fig. \ref{fig:smallturnaround}. Given the extraordinary similarity of the properties of 
curves in this figure for smallest values of departures between quantum and effective dynamics, and the behavior near turnaround in Fig. \ref{fig:largeturnaround} we expect that a non-monotonic behavior also exists 
for small values of $\omega_2^*$.

\subsection{Hubble rates, deceleration parameter and shear scalar}
Some important quantities which capture the approach to classical singularities are the mean Hubble rate, expansion ($\theta$) and the shear $({\sigma^2})$
scalars.\footnote{For a recent phenomenological discussion of these parameters in Bianchi-I spacetime, see Ref. \cite{sharif}.} For the Bianchi-I spacetime these can be expressed in terms of the directional Hubble rates $H_i$ as,
\be\label{expansion}
\theta = 3 H = H_1 + H_2 + H_3
\ee
where $H$ is the mean Hubble rate, 
and 
\be\label{shear}
\sigma^2 = \frac{1}{3} \left((H_2 - H_1)^2 + (H_3 - H_2)^2 + (H_3 - H_1)^2 \right) ~.
\ee
Another interesting parameter is the deceleration parameter defined as
\be\label{decel}
q = - \frac{\ddot a a}{\dot a^2} = -\frac{\dot H}{H^2} - 1 ~.
\ee
Here $a$ denotes the mean scale factor $a = (a_1 a_2 a_3)^{1/3}$.

The directional Hubble rates can be obtained using the Hamilton's equations, by using their definitions $H_i = \dot a_i/a_i$ which 
using the relations between the scale factors $a_i$ and $v_i$ (see Sec. II) result in 
\be\label{H_1}
H_1 = \frac{1}{3} \left(\frac{\dot v_2}{v_2} + \frac{\dot v_3}{v_3} - \frac{\dot v_1}{v_1} \right), ~ H_2 = \frac{1}{3} \left(\frac{\dot 
v_3}{v_3} + \frac{\dot v_1}{v_1} - \frac{\dot v_2}{v_2} \right) ~~ \mathrm{and} ~~ H_3 = \frac{1}{3} \left(\frac{\dot v_1}{v_1} + 
\frac{\dot v_2}{v_2} - \frac{\dot v_3}{v_3} \right) ~.
\ee
Note that the above definitions of the directional Hubble rates, mean Hubble rate, expansion and shear scalar and the decelration parameter 
 are valid 
in classical theory as well as LQC. Using the effective Hamiltonian constraint, in LQC the directional Hubble rates can be 
obtained using eq.(\ref{dotv1_v1}) and 
the corresponding equations for $\dot v_2/v_2$ and $\dot v_3/v_3$.

\begin{figure}[tbh!]
 \includegraphics[width=0.48\textwidth]{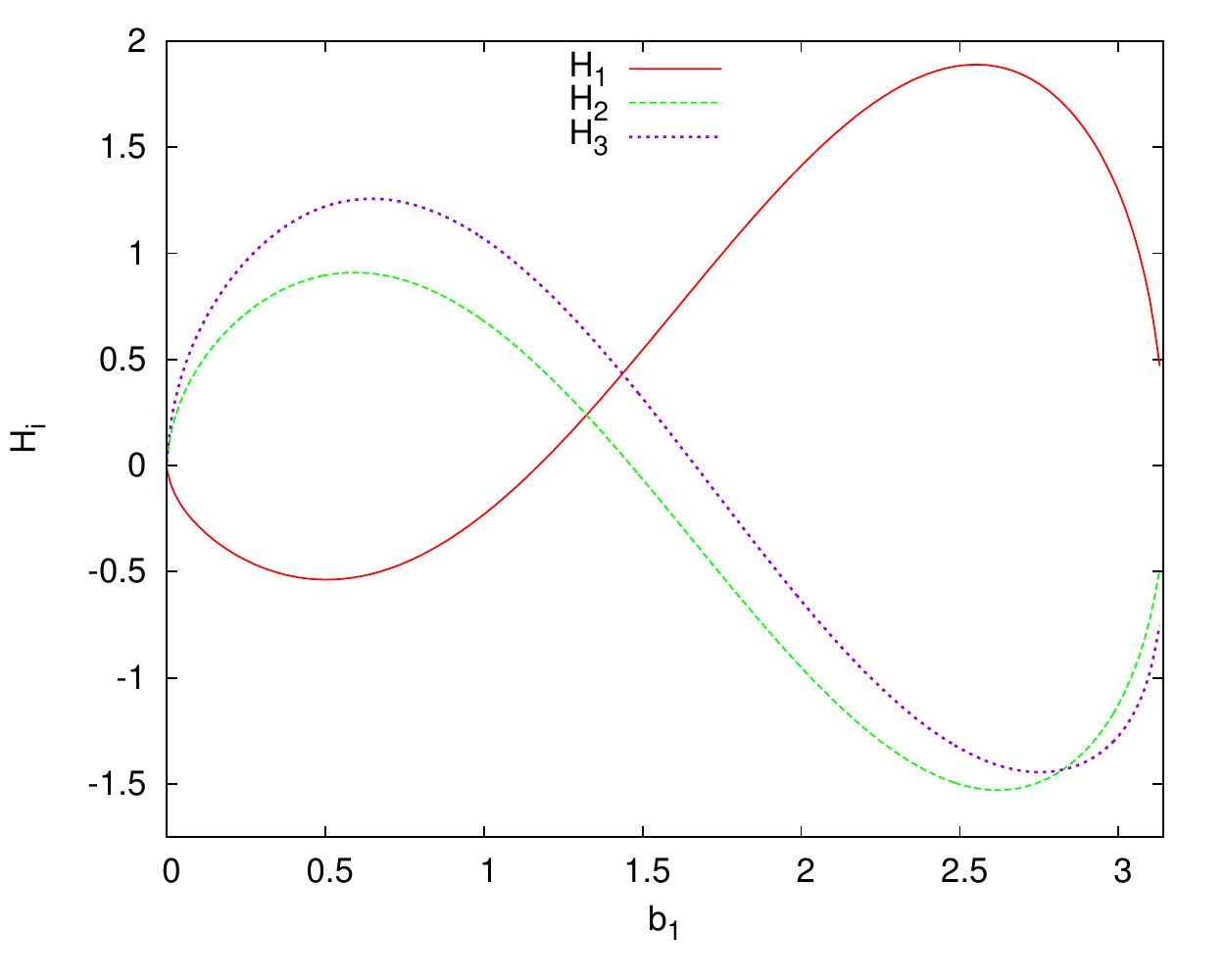}
 \caption{The behavior of directional Hubble rates (in Planck units) for a typical simulation is shown. For this simulation $\omega_2^*=750$ with $\omega_3^* 
= 1000$.} \label{fig:hubblerates}
\end{figure}

To estimate the mean Hubble rate (or equivalently the expansion scalar), shear scalar and deceleration parameter in the quantum theory, we assume that the states are sharply
peaked such that terms of the type $\langle \widehat{\cos(b_1)/V} \rangle$ in the operator version of (\ref{dotv1_v1}) can be approximated 
as $\langle \widehat{\cos(b_1)} \rangle/\langle \widehat{V} \rangle$. This approximation allows us to compute expectation values of mean Hubble rate 
(expansion scalar), shear scalar and the deceleration parameter in the quantum theory. It is important to note that this computation is different from all the other results 
shown in this section. In the previous results all the computed quantities were without such an approximation, and hence they were valid for 
all types of states.  On the other hand, the results for the mean Hubble rate (or the expansion scalar), the shear scalar and the decelration parameter require the above assumption, and usage of 
expectation values of the relevant operators in the definitions of mean Hubble rate (expansion scalar), shear scalar, and deceleration parameter (eqs.(\ref{expansion}), (\ref{shear}) and (\ref{decel})), 
with directional Hubble rates given by (\ref{H_1}). Unlike the results so far in this manuscript, the effective 
Hamiltonian constraint is used along with quantum expectation values to estimate the behavior of above quantities. 

\begin{figure}[!b]
 \includegraphics[width=0.48\textwidth]{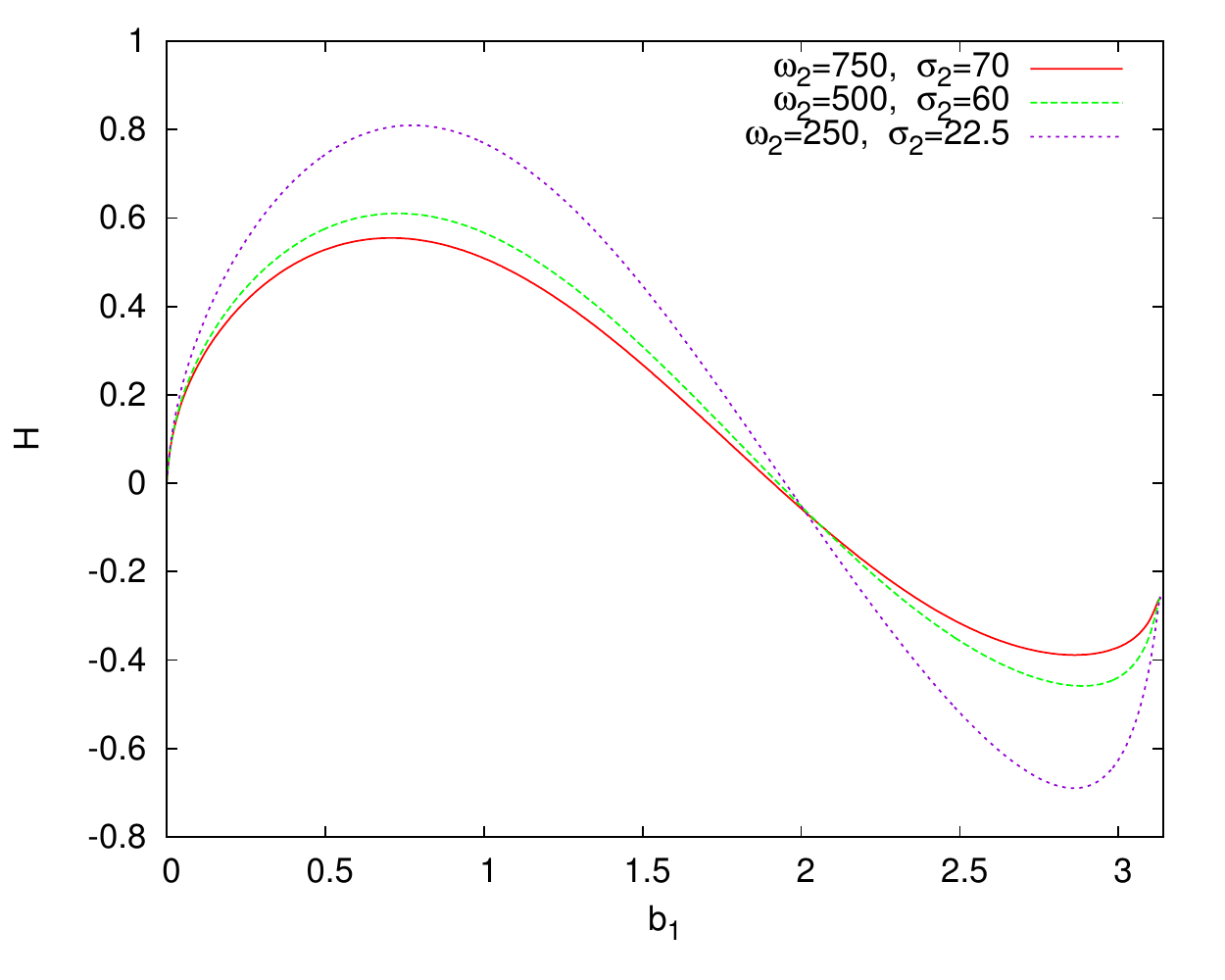}
 \caption{The mean Hubble rate $H$ (equal to $\theta$/3)  in Planck units is plotted versus $b_1$ for different values of $\omega_2^*$ with $\omega_3^* = 1000$.} \label{fig:expansion}
\end{figure}

\begin{figure}[t!]
 \includegraphics[width=0.48\textwidth]{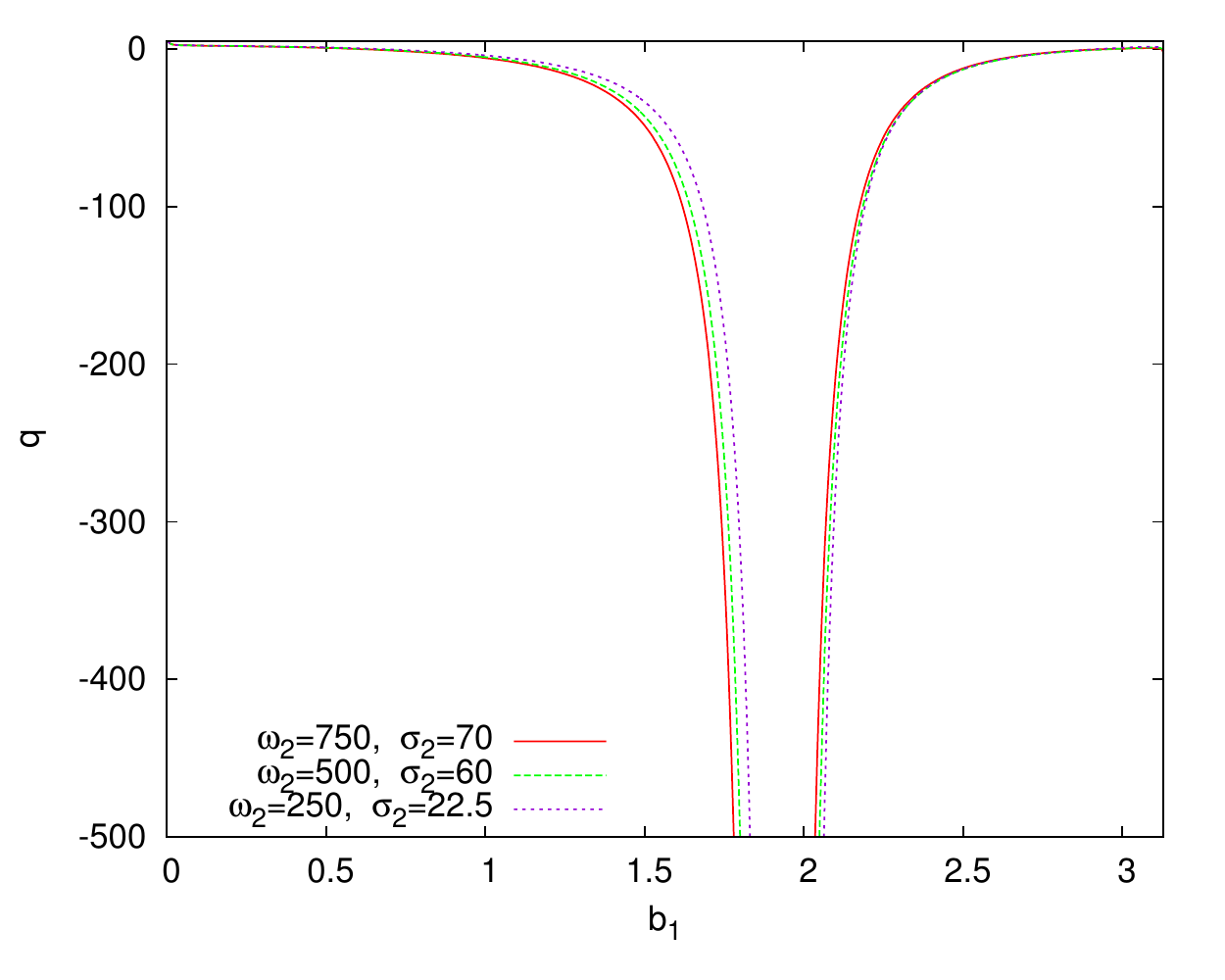}
 \caption{The deceleration parameter $q$ is plotted versus $b_1$ for different values of $\omega_2^*$ with $\omega_3^* = 1000$.} \label{fig:decel}
\end{figure}

The resulting behavior of the directional Hubble rates is shown in Fig. \ref{fig:hubblerates}. We have plotted the directional Hubble rates for a 
simulation for a Gaussian initial state peaked at $\omega_2^* = 750$ with $\sigma_2^* = 70$, with the values of $\omega_3^*$ and $\sigma_3$ 
 same as in all other simulations. The evolution of the Hubble rates 
confirm the anisotropic evolution. In this particular simulation, two directional Hubble rates $H_2$ and $H_3$ start with positive values 
and turn around to become negative at different times in $b_1$. The third Hubble rate $H_1$ start with a negative value and turns around to 
become positive. It is important to note that unlike the classical theory, Hubble rates remain bounded throughout the evolution. For a 
different choices of initial data, we obtain a similar behavior. Note that on changing the values of $\omega_2^*$ and $\omega_3^*$ in the 
initial state,  the  turnaround behavior and the maxima and minima of the directional Hubble rates changes. It should be further noted that at no value of 
internal time $b_1$ do the different Hubble rates ever coincide in an anisotropic evolution. This result is the consequence of Hamiltonian constraint (\ref{effham}) whose 
satisfaction at all times rules out 
all of the directional Hubble rates ever becoming equal.\footnote{A straightforward way to see this that the quantum constraint at large scales is a sum of terms $H_1 H_2 + H_2 H_3 + H_3 H1$ which should vanish. In an anisotropic evolution, the directional Hubble rates 
thus can;t be equal.} For the anisotropic evolution at least 
two of the directional Hubble rates would have to be different. If all the directional Hubble rates become equal then it will imply that the anisotropic shear is identically zero at a 
finite time which is not possible for the vacuum Bianchi-I spacetime.

The mean Hubble rate $H$ for the above simulation, along with two more cases is shown in Fig. \ref{fig:expansion}. (This behavior equivalently captures the expansion scalar 
$\theta$). It remains 
bounded in the entire evolution with a turnaround from positive to negative values. The vanishing of the expansion scalar or the mean Hubble rate provides us a 
value of bounce time $b_1$ for the mean volume of this Bianchi-I spacetime. The behavior of the mean Hubble rate is similar for all the 
simulations shown in Fig. \ref{fig:expansion}, albeit with a notable difference that for a fixed value of $\omega_3^*$, smaller values of 
$\omega_2^*$ yield larger absolute values of the mean Hubble rate.  For different simulations, we obtain similar results with an observation 
that there is no universal bound on the mean Hubble rate for different initial data. It is to be noted that the behavior of the expansion 
scalar confirms the expectations from an earlier result \cite{cs-geom}. In the latter work it was shown that due to a specific form of a 
polymerization in the Hamiltonian constraint, the expansion scalar has no universal bound unlike the alternative loop quantization of 
Bianchi-I spacetime \cite{awe-bianchi1,cs-geom}. The effective theory predicts that for states which probe the classical big bang 
singularity extremely closely such that $v_i$ almost vanish, the value of expansion scalar can be very high \cite{cs-geom}. The same is 
true for the directional Hubble rates and the shear scalar, which are also not bounded universally in this quantization. 
To verify this,  we would need initial data with much smaller 
values of $\omega_i$'s than what we were able to consider in this manuscript. Such states would result in extremely large computational 
requirements. Nevertheless, we found that as $\omega_2^*$ is decreased, the maximum value of $|H|$ increases. For one of our 
simulations probing deep Planck regime corresponding to $\omega_2^* = 40$ if we assume the validity of effective Hamiltonian approach to obtain above estimate, then the 
maximum value of $|\theta|$ increases to about five in Planck units. 

\begin{figure}[t!]
 \includegraphics[width=0.48\textwidth]{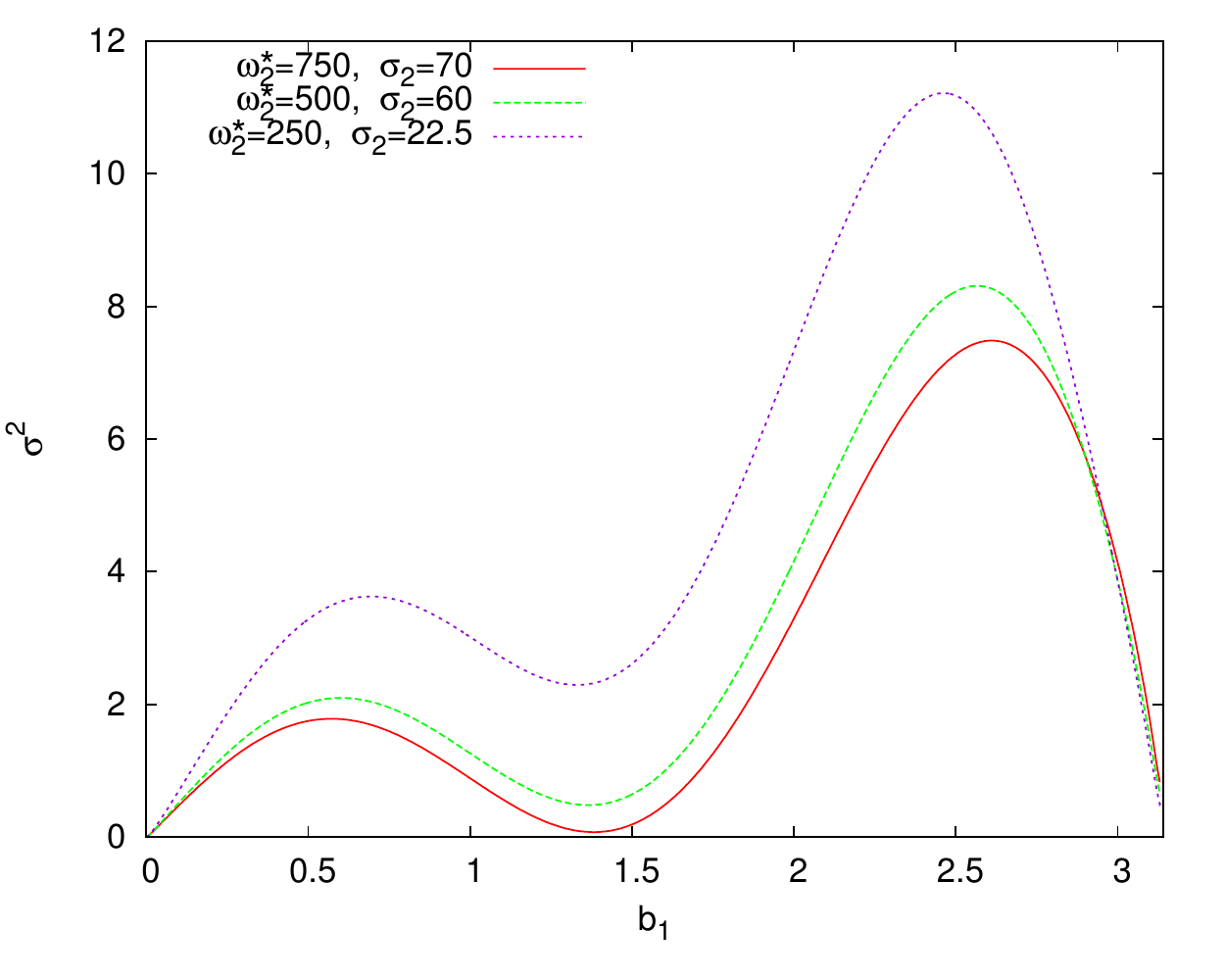}
 \caption{The behavior of the shear scalar (in Planck units) is shown in $b_1$ for various values of $\omega_2^*$ with a fixed value of $\omega_3^* = 1000$.} \label{fig:shear}
\end{figure}

The deceleration parameters for the simulations discussed in Fig. \ref{expansion} are plotted in Fig. \ref{fig:decel}. The deceleration parameter starts from negative values and its absolute  value increases 
towards the bounce of the mean scale factor. It's value reaches negative infinity when the mean Hubble rate vanishes. After the bounce of  the mean scale factor, deceleration 
parameter sharply increases though remains negative in the entire evolution. The behavior of the deceleration parameter turns out to be qualitatively similar for 
different values of $\omega_2^*$ as is evident from the above figure.

In Fig. \ref{fig:shear} we show the variation of shear scalar ($\sigma^2$) in time $b_1$ for three simulations corresponding to $\omega_2^* = 250$, $\omega_2^* = 500$ and 
$\omega_2^* = 750$. For these simulations, the shear scalar is bounded throughout the evolution. The same is true for all the simulations which we carried out in our analysis. 
This is in contrast to the classical theory where the shear scalar diverges signaling geodesic incompleteness and divergence in spacetime curvature. As in the case of the 
expansion scalar, we find that  smaller values of $\omega_2^*$ for a fixed $\omega_3^*$ result in a larger value of shear scalar. There is no universal bound, and for very small values of $\omega_2^*$, such as $\omega_2^* = 40$, we found that shear scalar can become larger than 30 in Planck units. Note that this is under the same assumptions of the validity of effective Hamiltonian which is found to be less reliable for small values of $\omega_2^*$.

\section{Discussion}

Let us begin with a summary of the main objectives and results of our analysis. In the last decade, loop quantization of various isotropic 
and anisotropic models has been performed. But only for the isotropic models, a robust picture of the singularity resolution and  Planck 
scale physics was so far available. In the seminal work of Refs. \cite{b1-madrid1,b1-madrid2}, analytical understanding of a loop 
quantization of vacuum Bianchi-I spacetime and singularity resolution using numerical methods was demonstrated. However, robustness of 
bounce and the associated physics for states with a wide variety of quantum fluctuations and the regime of validity of effective dynamics were  not 
investigated so far. As we discussed in Sec. III, the computational complexity and costs involved in performing simulations of 
anisotropic quantum spacetimes are enormously high in comparison to the isotropic models. To extract above physics an 
efficient parallelization of the codes and using HPC becomes essential. This constitutes the first main result of our paper. We have 
implemented the vacuum Bianchi-I spacetime in LQC in the {\code{Cactus}} framework which allows us to use conventional numerical relativity 
tools to perform numerical simulations of loop quantized spacetimes on HPC. The performance and efficiency of our implementation has 
been tested which shows excellent results. This technical feat allows us to extract the  physics of the deep quantum regime of loop 
quantized vacuum Bianchi-I spacetime. 

With over a hundred simulations performed for sharply peaked and 
widely spread Gaussian states we confirm that singularities are always resolved. Classical big bang singularity is replaced by the 
bounce(s) of directional volumes $v_i$. Each quantum state is characterized by $\omega_2$ and $\omega_3$ (along with their dispersions) 
which capture the anisotropy of the spacetime. We worked with the mixed representation with $b_1$ as relational time, and understood in detail the quantum expectation 
values of relational observables such as $\widehat{\ln (v_2)|_{b_1}}$. Keeping $\omega_3$ to be fixed, we varied $\omega_2$ to parameterize 
different initial states. In principle, one can vary  both but what matters in the anisotropic evolution is how one changes with respect to 
another. Varying just one of the two keeps dependence of results on the changes in $\omega_i$ transparent.  We find that for states which 
are peaked at large values of $\omega_2$, the effective dynamics is an excellent approximation to the underlying quantum dynamics. We find 
evidence of departure between the two as $\omega_2$ is decreased which also corresponds to bounce in directional volumes occurring at 
the lower values. The departure turns out to be most significant in the bounce regime and is thus measured as the difference between the 
logarithm of bounce volumes in $v_2$ in the quantum and effective dynamics. 

We quantified the departure of the effective dynamics from quantum dynamics in detail by  studying its 
dependence on the values of $\omega_2$ where the initial state is peaked, dispersions $\sigma_2$, relative dispersions of $\omega_2$ and 
finally the dispersions in the logarithm of the directional volume $v_2$. We find that keeping the dispersion in $\omega_2$ and other 
parameters fixed, if we decrease $\omega_2$ to smaller values then the departure of the effective dynamics from the quantum theory 
always increases at smallest values. However, for certain values of dispersion this behavior is not monotonic. For simulations 
corresponding to a large range of $\omega_2$, we found non-monotonicity before the above increase in departure occurs. This behavior is 
also captured if we study the dependence of departure between effective dynamics and quantum theory on dispersions in $\omega_2$. 
It shows that as the dispersion in $\omega_2$ decreases,  the departures first grow slowly but then increase very rapidly. Similar 
behavior is seen when dependence of the deviation between the effective and quantum dynamics is studied with respect to the relative 
dispersion in $\omega_2$. We find that there is a slow growth in departure for larger relative fluctuations which becomes very strong for 
smaller relative fluctuations. Larger values of $\omega_2$ result in pushing the phase of rapid increase to lower values in relative 
dispersion. The behavior of above departure with respect to dispersions in logarithm of directional volume $v_2$ brings forth an 
interesting non-monotonic behavior. The departure is largest as well as smallest for the large dispersions. There is a minimum 
allowed dispersion in $\omega_2$ and in logarithm of directional volume for any given value of $\omega_2$ on which an initial state is 
peaked. The difference between the quantum expectation value of the logarithm of directional volume $v_2$ and its counterpart computed at 
bounce is almost always positive. This implies that effective dynamics, in general underestimates the bounce volume an effect which 
becomes more pronounced for states with large dispersions in logarithm of volume (keeping other state parameters fixed). The same effect 
was earlier found in the numerical simulations of the homogeneous and isotropic models in LQC \cite{numlsu-2,numlsu-3}. However, we also 
find that for some simulations, corresponding to large dispersions in both directional volume and $\omega_2$, the above difference becomes 
negative. This means that at least in some cases, effective dynamics overestimates the bounce volume. It should be emphasized that these 
cases correspond to the computationally most involved and costly simulations and further work is needed to gain insights on this result. 
In summary, the behavior of the departure of effective dynamics from quantum theory shows an intricate and subtle relationship with the 
state parameters. Though these results establish the usefulness of effective dynamics for a large range of parameters, care must always be 
taken to generalize the results of effective dynamics arbitrarily. Especially the role of fluctuations is quite non-trivial and needs to be 
understood in detail analytically. 

In the classical theory, the approach to singularity is characterized by the divergences in the directional Hubble rates, and expansion and 
shear scalars becoming infinite. We found their behavior for the sharply peaked states. For these states  the effective 
Hamiltonian captures the quantum  dynamics quite faithfully. This computation used elements both from the quantum expectation 
values and the expressions of the above physical quantities found using effective Hamiltonian constraint. The behavior of directional 
Hubble rates, expansion and shear scalars turns out to be bounded throughout the evolution. However, it is not universally bounded by a 
particular value, as for example predicted in the effective dynamics of alternative loop quantization of Bianchi-I spacetime \cite{cs-geom}. 
We found that for certain simulations the expansion and shear scalars can be quite large than Planckian values. We also studied the behavior of the deceleration parameter 
in this spacetime. Since this parameter is inversely proportional to the square of the mean Hubble rate, it diverges at the bounce of the mean volume and remains finite in the 
other regime. We note that given the lack of matter fields in our fully quantized model, one can not use the construction as is for phenomenological studies and investigate 
other phenomenologically interesting parameters and effects of matter content such as dark energy at late times. In the classical theory, various such studies have been performed (see for eg. \cite{sharif}). For the latter kind of a study, it will be important to include matter in the Hamiltonian constraint at the quantum level, and perform numerical simulations 
with additional grids in the matter phase space variables. This inevitably increases the numerical demand of resources used in the simulations and is beyond the scope of the present analysis. This exercise, along with 
study of phenomenologically interesting parameters in relation to cosmological models will be performed else where. We emphasize that even for such matter models the approach to singularity is captured by Bianchi-I vacuum spacetime which 
has been quantized and rigorously studied numerically in the present manuscript. Thus, our analysis provides insights on the fate of singularities to various matter models in Bianchi-I spacetime using techniques of loop quantum gravity.

Let us comment on two interesting issues. The first one deals with the genericity of singularity resolution and bounce in this model. And the second one on how different or similar is the 
physics across the bounce. An important step to prove the first statement has been carried out in this work. Our results from this analysis show that the singularity resolution and bounce occur for a wide variety of Gaussian states. In an upcoming work this conclusion is extended to the 
squeezed and non-Gaussian states \cite{next}. Using the properties of the quantum Hamiltonian constraint, it can be shown that the state corresponding to the zero volume, 
where the classical singularity occurs, does not lie in the physcial Hilbert space in the loop quantization of this 
spacetime \cite{b1-madrid1}. In addition, it has been shown using the effective spacetime description of the Bianchi-I spacetime that bounce always occurs within the limit of validity of effective dynamics \cite{ps11}. 
Our current analysis strongly validates the applicability of effective dynamics and hence these results. All these results point to a strong evidence that singularity resolution always occurs in this particular model in LQC and 
bounce is a generic phenomena for a wide class of states. Now let us visit the second issue. Once we have the evidence of genericness of bounce in what sense the spacetime before and after the bounce is similar and different? 
The answer depends on the what features and variables we are interested in to extract physics. It turns out that the shear scalar though varies a lot during the bounce phase, takes the same value at very early times before 
the bounce of the mean volume and at very late times after the bounce \cite{kevin-chiou}. This is confirmed from Fig. \ref{fig:shear} by comparing values of the shear scalar near $b_1 = 0$ (late times after the bounce) and 
$b_1 = \pi$ (early times before the bounce). This implies that the anisotropic shear is preserved across the bounce. On the other hand the geometrical structure of spacetime as the classical singularity is approached can be quite 
different before and after the bounce, esepcially if matter is present \cite{bgps-kasner}. In the present case of the Bianchi-I vacuum spacetime, this structure is cigar type before and after the bounce (as is evident 
from Fig. \ref{fig:hubblerates}). From the behavior of different variables we studied in our analysis, the physics of the loop quantized Bianchi-I vacuum spacetime turns out to be similar before and after the bounce. This can though change 
in matter models for this spacetime, for example in the presence of inflation in loop quantum Bianchi-I spacetime \cite{bgps-inflation}.

Results obtained in this manuscript open a new avenue to understand the detailed nature of anisotropic spacetimes in LQC in the full quantum theory. With the tools we 
have introduced, numerical simulations of Bianchi-II and Bianchi-IX spacetimes as well as Kantowski-Sachs model can become feasible. Apart from these, 
an important step in this direction will be to perform numerical simulations for the other loop quantization of Bianchi-I spacetime as 
proposed in Refs. \cite{chiou-bianchi1,awe-bianchi2}, and compare with the results in the current analysis. Recall that this quantization 
prescription does not require to restrict the spatial manifold to be a 3-torus. In fact, using effective dynamics for this quantization 
generic resolution of all the strong singularities is predicted \cite{ps11}. As we mentioned earlier, a better analytical understanding of 
the physical Hilbert space is needed to perform numerical analysis for this quantization. Some of the properties of the quantum Hamiltonian 
constraint were understood in Ref. \cite{kp-fluc}, and recently some other analytical challenges to understand the physical Hilbert space 
have been overcome \cite{tomek}. Despite this progress, numerical analysis faces a further complexity because the quantum difference equation 
for the quantization in Refs. \cite{chiou-bianchi1,awe-bianchi1} has an additional non-locality. Perhaps using methods proposed in Ref. 
\cite{sabharwal-khanna} in synergy with techniques used in this paper would allow numerical simulations for the alternative loop 
quantization of the Bianchi-I spacetime.

We now comment on some of the questions in the present quantization of Bianchi-I model which can be answered in future works using 
techniques built in this analysis. Our work dealt with the study of sharply peaked as well as widely spread Gaussian states. While this has 
given useful insights on the robustness of singularity resolution and the validity of effective dynamics, it is nevertheless important to 
extend these results to more general states such as squeezed and non-Gaussian states. Such a study will provide  answers to whether 
singularity resolution in anisotropic spacetime occurs for arbitrary states.  Computationally this would bring additional challenges and 
generalization of the Chimera scheme \cite{numlsu-1} to anisotropic spacetimes would be essential. Another direction which is essential to 
be explored is understanding bounds on growth of fluctuations through bounces. In all the simulations considered in our analysis, we found 
that 
fluctuations of the initial states never grew unbounded in the evolution. Rather these fluctuations  
are seemingly preserved across the 
bounce. The situation mimics the case 
of isotropic models where analytical \cite{cs-recall,cm-fluc1,*cm-fluc2,kp-fluc} and numerical 
results \cite{aps2,aps3,apsv,numlsu-2,numlsu-3} on constraints on the growth of fluctuations in LQC are available. 
Though, analytical results 
are available for the alternate prescription of loop quantization of Bianchi-I spacetime \cite{kp-fluc}, they need to be extended to include 
the present quantization. Numerical analysis presented in this manuscript can be used to understand constraints on the growth of the 
fluctuations through anisotropic bounces. Finally, it is crucial to employ new tools for a better visualization and extraction of the 
multi-variate data from simulations. The limitation with 
conventional ways is apparent by noting that one can only view and analyze only a few correlations at once. 
While this does not pose an issue if one is interested in understanding the behavior of a few relational observables, the richness of 
anisotropic models can not be fully appreciated. Addressing this issue will allow understanding correlations and dependence of many 
physical quantities at once, giving deeper insights in to the physics of quantum anisotropic and black hole spacetimes.

\acknowledgements
We are grateful to Brajesh Gupt for various discussions during the initial stages of this work. This work is
supported by NSF grants PHY-1404240 and PHY-1454832.  This work is also supported by a grant from John
Templeton Foundation.  The opinions expressed in this publication are those of authors and do not
necessarily reflect the views of John Templeton Foundation. This work used the Extreme Science and Engineering Discovery 
Environment (XSEDE), which is supported by National Science Foundation grant number ACI-1053575.

\section*{References}

\end{document}